# Quantization of magnetoelectric fields

E. O. Kamenetskii

Microwave Magnetic Laboratory,
Department of Electrical and Computer Engineering,
Ben Gurion University of the Negev, Beer Sheva, Israel

January 22, 2018

**Abstract**
The effect of quantum coherence involving macroscopic degree of freedom, and occurring in systems far larger than individual atoms are one of the topical fields in modern physics. Because of material dispersion, a phenomenological approach to macroscopic quantum electrodynamics, where no canonical formulation is attempted, is used. The problem becomes more complicated when geometrical forms of a material structure have to be taken into consideration. Magnetic-dipolar-mode (MDM) oscillations in a magnetically saturated quasi-2D ferrite disk are macroscopically quantized states. In this ferrimagnetic structure, long-range dipole-dipole correlation in positions of electron spins can be treated in terms of collective excitations of a system as a whole. The near fields in the proximity of a MDM ferrite disk have space and time symmetry breakings. Such MDM-originated fields – called magnetoelectric (ME) fields – carry both spin and orbital angular momentums. By virtue of unique topology, ME fields are different from free-space electromagnetic (EM) fields. The ME fields are quantum fluctuations in vacuum. We call these quantized states ME photons. There are not virtual EM photons. We show that energy, spin and orbital angular momenta of MDM oscillations constitute the key physical quantities that characterize the ME-field configurations. We show that vacuum can induce a Casimir torque between a MDM ferrite disk, metal walls, and dielectric samples.

## I. INTRODUCTION

Energy, linear momentum, and angular momentum are constants of motion, which constitute the key physical quantities that characterize the EM field configuration. Their conservation can be cast as a continuity equation relating to a density or a flux density, or a current, associated to the conserved quantity [1]. Recently, a new constant of motion and a current associated to this conserved quantity have been introduced in electrodynamics. For a circularly polarized EM plane wave, there are the EM-field chirality and the flux density of the EM-field chirality [2, 3].

Circularly polarized EM photons carry a spin angular momentum. EM photons can also carry an additional angular momentum, called an orbital angular momentum. Azimuthal dependence of beam phase results in a helical wavefront. The photons, carrying both spin and orbital angular momentums are twisted photons [4]. The twisted EM photons are propagating-wave, "actual", photons. In the near-field phenomena, which are characterized by subwavelength effects and do not radiate though space with the same range properties as do EM wave photons, the energy is carried by virtual EM photons. Virtual particles should also conserve energy and momentum. The question whether virtual EM photons can behave as twisting excitations is a subject of a strong interest. Recently, a new concept of the spin-orbit interactions in the evanescent-field region of optical EM waves has been proposed.

Theoretically, it was shown that an evanescent wave possesses a spin component, which is orthogonal to the wave vector. Furthermore, such a wave carries a momentum component, which is determined by the circular polarization and is also orthogonal to the wave vector. The transverse momentum and spin push and twist a probe Mie particle in an evanescent field. This should allow the observation of 'impossible' properties of light, which was previously considered as 'virtual' [5, 6].

The evanescent modes have a purely imaginary wave number and represent the mathematical analogy of the tunneling solutions of the Schrödinger equation. However, the quantization of evanescent waves is not a straightforward problem. Many fundamental optical properties rely on virtual photons to act as the mediator. A well-known example of the Casimir effect can be understood as involving the creation of short lived virtual photons from the vacuum. Together with the calculation of the Casimir force between arbitrary materials, in numerous publications, the question of the angular-momentum coupling with the quantum vacuum fields is considered as a topical subject. Since photons also carry the angular momentum, the vacuum torque will appear between macroscopic bodies when their characteristic properties are anisotropic [7 − 13]. In particular, in Refs. [7, 8, 10, 13], it was discussed that vacuum can induce a torque between two axial birefringent dielectric plates. In this case, the fluctuating electromagnetic fields have boundary conditions that depend on the relative orientation of optical axes of the materials. Hence, the zero-point energy arising from these fields also has an angular dependence. This leads to a Casimir torque that tends to align two of the principal axes of the material in order to minimize the system's energy. A torque occurs only if symmetry between the right-handed and left-handed circularly polarized light is broken (when the media are birefringent).

The contributions of plasmonic modes to the Casimir effect was analyzed in recent studies [14 − 17]. Plasmon polaritons are coupled states between electromagnetic radiation and electrostatic (ES) oscillations in a metal. Instantaneous Coulomb interaction are supplemented with retardation effects associated with 'real' electromagnetic fields. In a non-classical description, there is the effect of interaction between real and virtual photons. On the other hand, recently, it was shown that some special-geometric plasmonic structures can be used to create so called superchiral fields [18, 19]. Could one consider these near fields as chiral virtual photons? Whether the contribution of such chiral virtual photons to the Casimir torque can be analyzed? To the best of our knowledge, no such effects of are discussed in literature.

Magnetostatic (MS) oscillations in ferrite samples [20, 21] can be considered, to a certain extent, as a microwave analog of plasmon oscillations in optics. The coupled states between electromagnetic radiation and MS oscillations may be viewed as a microwave MS-magnon polaritons, analogously to the ES-plasmon polaritons in optics. Such an analogy, however, is quite inaccurate, since dynamics of charge motion is completely different in these cases. In contrast to ES-plasmon oscillations with linear electric currents, MS-magnon oscillations appear due to precessing magnetic dipoles [20 − 22]. The problem of MS [or, in other words, magnetic-dipolar-mode (MDM)] oscillations demonstrates fundamentally new properties when specific geometrical forms of a material structure is taken into consideration. In a quasi-2D ferrite disk, one becomes evident with a fact that MDM oscillations are macroscopically (mesoscopically) quantized states.

The effects of quantum coherence involving macroscopic degrees of freedom and occurring in systems far larger than individual atoms are one of the topical fields in modern physics [23]. There is evidence that macroscopic systems can under appropriate conditions be in quantum states, which are linear superpositions of states with different macroscopic properties. The Aharonov-Bohm effect shows that the characteristically quantum effect is reflected in the behavior of macroscopic currents [24]. Much progress has been made in demonstrating the macroscopic quantum behavior of superconductor systems, where particles form highly correlated electron systems. The concept of coherent mixture of electrons and holes, underlying the quasiclassical approximation based on the Bogoliubov-de Gennes (BdG) Hamiltonian [25], well describes the topological superconductors.

Macroscopic quantum coherence can also be observed in some ferrimagnetic structures. Long range dipole-dipole correlation in position of electron spins in a ferrimagnetic sample with saturation magnetization can be treated in terms of collective excitations of the system as a whole. If the sample is sufficiently small so that the dephasing length $L_{ph}$ of the magnetic dipole-dipole interaction exceeds the sample size, this interaction is non-local on the scale of $L_{ph}$. This is a feature of a mesoscopic ferrite sample, i. e., a sample with linear dimensions smaller than $L_{ph}$ but still much larger than the exchange-interaction scales. In a case of a quasi-2D ferrite disk, the quantized forms of these collective matter oscillations – MS (or MDM) magnons – were found to be quasiparticles with both wave-like and particle-like behaviors, as expected for quantum excitations. The MDM oscillations in a quas-2D ferrite disk, analyzed as spectral solutions for the MS-potential scalar wave function $\psi(\vec{r},t)$, has evident quantum-like attributes. The oscillating state in a lossless ferrite-disk particle can be described in terms of a Hilbert space spanned by a complete orthonormal set of eigenvectors of some observable $A$ with eigenvalues $a_i$ [26 – 28]. The discrete energy eigenstates of the MDM oscillations are well observed in microwave experiments [29 – 32]. Experimentally, it was shown also that interaction of the MDM ferrite particles with its microwave environment give multiresonance Fano-type resonances [31, 32]. The observed interference between resonant and nonresonant processes is typical for quantum-dot structures [33]. In this case, the Fano regime emerges because of resonant tunneling between the dot and the channel.

MDM oscillations in a quasi-2D ferrite disk can conserve energy and angular momentum. Because of these properties, MDMs strongly confine energy in subwavelength scales of microwave radiation. In a vacuum subwavelength region abutting to a MDM ferrite disk, one observes the quantized state of power-flow vortices. Moreover, in such a vacuum subwavelength region, the time-varying electric and magnetic fields can be not mutually perpendicular. Such a specific near field – so-called magnetoelectric (ME) field – give evidence for spontaneous symmetry breakings at the resonance states of MDM oscillations [34, 35]. The ME fields are quantum fluctuations in vacuum. We call them ME photons. The electric- and magnetic-field components of the ME photons have spin and orbital angular momenta. These twisted evanescent fields are neither virtual nor "real" (free-space propagating) EM photons in vacuum.

The fact of the presence of the power flow circulation inside and in a subwavelength vacuum region outside a ferrite disk arises a question of the angular momentum balance. It is evident that at the MDM resonance, such a balance can be realized only when the opposite power flow circulations can be created due to metal (and also dielectric) parts of the microwave structure, in which a ferrite disk is placed. It means that at the MDM resonance in a ferrite disk, certain conductivity (or displacement) electric currents should be induced in the metal (or dielectric) parts of the microwave structure. The coupling between an electrically neutral MDM ferrite disk and an electrically neutral metal (or dielectric) objects in a microwave subwavelength region, should involve the electromagnetic-induction or/and Coulomb interactions through virtual photons mediation. In Ref. [36], it was shown that due to the topological action of the azimuthally unidirectional transport of energy in a MDM-resonance ferrite sample there exists the opposite topological reaction on a metal screen placed near this sample. We call this effect topological Lenz's effect. The coupling energy depends on the angle between the directions of the magnetization vectors in a ferrite and electric-current vectors on a metal wall. A vacuum-induced Casimir torque [7 – 13] allows for torque transmission between the ferrite disk and metal wall avoiding any direct contact between them.

In this paper, we analyze quantum confinement of MDM oscillations in a ferrite disk and coupling between MDM bound states and microwave-field continuum. The paper is organized as follows. In section II, we consider MDM oscillations and ME fields based on a classical approach. We analyze the MS description in connection with the Faraday law and "Sagnac effect" in a MDM ferrite disk. We show that the near fields originated from magnetization dynamics at MDM resonances – the ME fields – appear as the fields of axion electrodynamics. Sharp multiresonance oscillations, observed experimentally in microwave structures with an embedded quasi-2D ferrite disk give evidence for quantized states of the microwave fields originated from MDM oscillations. This is a subject of our detailed discussion in section III. In section IV, we consider MDM bound states in the waveguide continuum. The questions of *PT* symmetry and *PT*-symmetry breaking of MDM oscillations are the subject of section V. In section VI, we analyze the effects in non-Hermitian MDM structures with *PT*-symmetry breaking. Section VII is a conclusion of the paper.

## II. MDM OSCILLATIONS AND MAGNETOELECTRIC FIELDS: CLASSICAL APPROACH

### A. Magnetostatic description and Faraday law

In a frame of a classical description, MDM oscillations in small ferrite samples are considered as an approximation to Maxwell equations when a displacement electric current is negligibly small. The physical justification for such an approximation arises from the fact that in a small (with sizes much less than a free-space electromagnetic wavelength) sample of a magnetic material with strong temporal dispersion (due to the ferromagnetic resonance), one neglects a time variation of electric energy in comparison with a time variation of magnetic energy [20 – 22]. In this case, we have a system of three differential equations for the electric and magnetic fields

$$\nabla \cdot \vec{B} = 0, \quad (1)$$

$$\vec{\nabla} \times \vec{H} = 0, \quad (2)$$

$$\vec{\nabla} \times \vec{E} = -\frac{\partial \vec{B}}{\partial t}. \quad (3)$$

In such a system, there is evident electromagnetic duality breaking. The spectral solutions for MDM oscillations in a small ferrite sample can be obtained based on the first two differential equations in the system, Eqs. (1) and (2). With a formal use of equation $\vec{H} = -\vec{\nabla}\psi$ and a constitutive relation for a ferrite, one gets the Walker equation [37]

$$\vec{\nabla} \cdot \left( \ddot{\mu} \cdot \vec{\nabla} \psi \right) = 0 \quad (4)$$

inside and the Laplace equation $\nabla^2 \psi = 0$ outside a ferrite sample. Here $\psi$ is a magnetostatic (MS) potential and $\ddot{\mu}(\omega, \vec{H}_0)$ is a tensor of ferrite permeability at the ferromagnetic-resonance frequency range.

To obtain the MDM spectral solutions, the boundary conditions for the MS-potential scalar wave function $\psi(\vec{r},t)$ and its space derivatives should be imposed. In these spectral solutions, we do not use the third equation in the system – the Faraday equation (3). From literature, one can see that for the MDM spectral problem formulated exceptionally for the MS-potential wave function $\psi(\vec{r},t)$, no use of the alternative electric fields is presumed [20 – 22, 37 – 39]. Following a formal analysis [40, 41] it can be also shown that for MDM oscillations, the Faraday equation is incompatible with Eqs. (1) and (2). Really, from Eq. (3), we obtain $\vec{\nabla} \times \frac{\partial \vec{E}}{\partial t} = -\frac{\partial^2 \vec{B}}{\partial t^2}$. Excluding completely the electric displacement current $\left( \frac{\partial \vec{D}}{\partial t} = 0 \right)$, in a sample which does not possess any dielectric anisotropy, we have $\frac{\partial^2 \vec{B}}{\partial t^2} = 0$. It follows that the magnetic field in small resonant magnetic objects varies linearly with time. This gives, however, arbitrary large fields at early and late times which is excluded on physical grounds. An evident conclusion suggests itself at once: at the MDM resonances, the magnetic fields are constant quantities. Since such a conclusion contradicts the fact of temporally dispersive media and any resonant conditions, one can state that the Faraday equation is incompatible with the MDM spectral solutions. In Ref. [42] it was merely stated that in a case of the MDM spectral problem, an electric field is completely absent and thus $\vec{\nabla} \times \vec{E} = 0$.

**B. "Sagnac effect" in a MDM ferrite disk**

It appears, however, that in a case of MDM oscillations in a quasi-2D ferrite disk, the Faraday equation plays an essential role. These oscillations are frequency steady states characterizing by power-flow azimuthal rotations. In a frame of reference co-moving with the orbital rotation, the fields are constant. In every steady state, the Faraday law appears as the Ampere-law analog for magnetic currents as sources of the electric fields. The following consideration makes this question clearer.

An analytical approach for MDM resonances in a quasi-2D ferrite disk is based on a formulation of a spectral problem for a macroscopic scalar wave function – the MS-potential wave function $\psi$ [26, 28, 34]. In this approach, the description rests on two cornerstones: (i) Precession of all electrons in a magnetically ordered ferrite sample is determines by $\psi$ function, and (ii) the phase of this wave function is well defined over the whole ferrite-disk system, i.e., MDMs are macroscopic states maintaining the global phase coherence. For a ferrite magnetized along $z$ axis, the permeability tensor has a form [20]:

$$\ddot{\mu} = \mu_0 \begin{bmatrix} \mu & i\mu_a & 0 \\ -i\mu_a & \mu & 0 \\ 0 & 0 & 1 \end{bmatrix}. \qquad (5)$$

In a quasi-2D ferrite disk with the disk axis oriented along $z$, the Walker-equation solution for the MS-potential wave function is written in a cylindrical coordinate system as [26, 28, 34, 35]:

$$\psi = C\xi(z)\tilde{\varphi}(r,\theta), \qquad (6)$$

where $\tilde{\varphi}$ is a dimensionless membrane function, $r$ and $\theta$ are in-plane coordinates, $\xi(z)$ is a dimensionless function of the MS-potential distribution along $z$ axis, and $C$ is a dimensional amplitude coefficient. Being the energy-eigenstate oscillations, the MDMs in a ferrite disk are also characterized by topologically distinct structures of the fields. This becomes evident from the boundary condition on a lateral surface of a ferrite disk of radius $\mathcal{R}$, written for a membrane wave function as [26, 28, 34, 35]:

$$\mu \left( \frac{\partial \tilde{\varphi}}{\partial r} \right)_{r=\mathcal{R}^-} - \left( \frac{\partial \tilde{\varphi}}{\partial r} \right)_{r=\mathcal{R}^+} + i \frac{\mu_a}{\mathcal{R}} \left( \frac{\partial \tilde{\varphi}}{\partial \theta} \right)_{r=\mathcal{R}^-} = 0. \tag{7}$$

Evidently, in the solutions, one can distinguish the time direction (given by the direction of the magnetization precession and correlated with a sign of $\mu_a$) and the azimuth rotation direction (given by a sign of $\frac{\partial \tilde{\varphi}}{\partial \theta}$). For a given sign of a parameter $\mu_a$, there can be different MS-potential wave functions, $\tilde{\varphi}^{(+)}$ and $\tilde{\varphi}^{(-)}$, corresponding to the positive and negative directions of the phase variations with respect to a given direction of azimuth coordinates, when $0 \leq \theta \leq 2\pi$. So a function $\tilde{\varphi}$ is not a single-valued function. It changes a sign when angle $\theta$ is turned on $2\pi$.

Inside a ferrite disk, the boundary-value-problem solution for Eq. (6) is written as

$$\psi(r,\theta,z,t) = C_{\nu,n} J_\nu \left( \frac{\beta r}{\sqrt{-\mu}} \right) \left( \cos \beta z + \frac{1}{\sqrt{-\mu}} \sin \beta z \right) e^{-i\nu\theta} e^{i\omega t}. \tag{8}$$

Here $\beta$ is a wave number of a MS wave propagating in a ferrite along the $z$ axis, $\nu$ is a positive integer azimuth number, and $J_\nu$ is the Bessel function of order $\nu$ for a real argument. This equation shows that the modes in a ferrite disk are MS waves standing along the $z$ axis and propagating along an azimuth coordinate in a certain (given by a direction of a normal bias magnetic field) azimuth direction. When the spectral problem for the MS-potential scalar wave function $\psi(\vec{r},t)$ is solved, distribution of magnetization in a ferrite disk is found as $\vec{m} = -\overset{\leftrightarrow}{\chi} \cdot \vec{\nabla} \psi$, where $\overset{\leftrightarrow}{\chi}$ is the susceptibility tensor of a ferrite [20, 21]. The magnetization has both the spin and orbital rotation. There is the spin-orbit interaction between these angular momenta. The electric field in any point inside or outside a ferrite disk is defined as [35, 43]

$$\vec{E}(\vec{r}) = -\frac{1}{4\pi} \int_V \frac{\vec{j}^{(m)}(\vec{r}') \times (\vec{r} - \vec{r}')}{|\vec{r} - \vec{r}'|^3} dV', \tag{9}$$

where $\vec{j}^{(m)} = i\omega\mu_0 \vec{m}$ is the density of a magnetic current. In Eq. (9), the frequency $\omega$ is a discrete quantity of the MDM-resonance frequency. The magnetic field inside a ferrite disk is easily defined from the equation $\vec{H} = -\vec{\nabla}\psi$. Based on the known magnetization $\vec{m}$ inside a ferrite, one can find also the magnetic field distribution at any point outside a ferrite disk [1, 43]:

$$\vec{H}(\vec{r}) = \frac{1}{4\pi} \left( \int_V \frac{(\vec{\nabla}' \cdot \vec{m}(r'))(\vec{r} - \vec{r}')}{|\vec{r} - \vec{r}'|^3} dV' - \int_S \frac{(\vec{n}' \cdot \vec{m}(r'))(\vec{r} - \vec{r}')}{|\vec{r} - \vec{r}'|^3} dS' \right). \tag{10}$$

In Eqs. (9) and (10), $V$ and $S$ are a volume and a surface of a ferrite sample, respectively. Vector $\vec{n}'$ is the outwardly directed normal to surface $S$.

The azimuthally unidirectional wave propagation of MDMs are observed in mechanically non-rotating quasi-2D ferrite disks. It is worth comparing these oscillations with Sagnac-effect resonances in microcavities. The Sagnac effect is manifested when a sample body rotates mechanically. In the case of the 2D disk dielectric resonator rotating mechanically at angular velocity $\Omega$, the resonances are obtained by solving the stationary wave equation [44]:

$$\frac{\partial^2 \varphi}{\partial r^2} + \frac{1}{r}\frac{\partial \varphi}{\partial r} + \frac{1}{r^2}\frac{\partial^2 \varphi}{\partial \theta^2} + 2ik\frac{\Omega}{c}\frac{\partial \varphi}{\partial \theta} + n^2 k^2 \varphi = 0, \qquad (11)$$

where $\varphi$ is an electric field component. The solutions for $\varphi(r,\theta)$ is given as $\varphi(r,\theta) = f(r)e^{im\theta}$, where $m$ is an integer. When the disk cavity is rotating, the wave function is the rotating wave $J_m(K_m r)e^{im\theta}$, where $K_m^2 = n^2 k^2 - 2k(\Omega/c)m$. The frequency difference between the counter-propagating waves is equal to $\Delta\omega = 2(m/n^2)\Omega$. For a given direction of rotation, the clockwise (CW) and counterclockwise (CCW) waves inside a dielectric cavity experience different refraction index $n$. Their azimuthal numbers are $\pm m$.

The wave equation for MDM oscillations in a quasi-2D ferrite disk, are similar, to a certain extent, to the solutions describing Sagnac effect in mechanically rotating optical microcavities. When we compare Eq. (7) with Eq. (11) we see that in both cases, there are the terms with the first-order derivative of the wave function with respect to the azimuth coordinate. Similar to the Sagnac effect in optical microcavities, MDM oscillations in a ferrite disk are described by the Bessel-function azimuthally rotating MS-potential waves [26]. However, in our case of MDMs, the field rotation is due to topological chiral currents on a lateral surface of a mechanically stable ferrite disk. These edge magnetic currents appear because of the spin-orbit interaction in magnetization dynamics.

The above discussion and statement that the Faraday equation is incompatible with the MDM spectral solutions we can clarify now for the case of a quasi-2D ferrite disk. It is known that in macroscopic electrodynamics, one can define three types of currents: density of the electric displacement current $\varepsilon_0 \frac{\partial \vec{E}}{\partial t}$, the electric current density arising from polarization $\frac{\partial \vec{p}}{\partial t}$, and the electric current density arising from magnetization $\vec{\nabla} \times \vec{m}$ [1, 22]. For a magnetic insulator [such as yttrium iron garnet (YIG)], we have the macroscopic Maxwell equation:

$$\vec{\nabla} \times \vec{B} = \mu_0 \left( \varepsilon_0 \frac{\partial \vec{E}}{\partial t} + \frac{\partial \vec{p}}{\partial t} + \vec{\nabla} \times \vec{m} \right), \qquad (12)$$

At the MDM resonances in YIG, orbitally rotating MS-potential waves $\psi(\vec{r},t)$ provide magnetization ($\vec{m} = -\vec{\vec{\chi}} \cdot \vec{\nabla}\psi$) vectors with spin and orbital rotation and, as a result, the $\vec{E}$ and $\vec{p}$ vectors with spin and orbital rotation. We have the situation when the lines of the electric field $\vec{E}$ as well the lines of the polarization $\vec{p}$ are "frozen" in the lines of magnetization $\vec{m}$. It means that there are no time variations of vectors $\vec{E}$ and $\vec{p}$ with respect to vector $\vec{m}$ (and, certainly, with respect to space derivatives of vector $\vec{m}$). In other words, all the time variations of vectors $\vec{E}$ and $\vec{p}$ are synchronized with the time variations of vector $\vec{m}$.

In other words, the lines of the electric field $\vec{E}$ as well the lines of the polarization $\vec{p}$ in a sample are "frozen" in the lines of magnetization $\vec{m}$. At this assumption, with taking into account the constitutive relation $\vec{B} = \mu_0 (\vec{H} + \vec{m})$, Eq. (12) is definitely reduced to Eq. (2). In such a case, the Faraday law (3) is not in contradiction with the MS equations (1) and (2). We can say that the entire system of Eqs. (1) – (3) is correct in the coordinate frame of *orbitally driven field patterns*.

At the MDM resonance, the electric field originated from a ferrite disk can cause electric polarization $\vec{p}$ in a dielectric sample situated outside a ferrite. In the reference frame co-rotating with the magnetization in a ferrite disk, this electric polarization in a dielectric sample is not time varying. In the orbitally rotating field patterns, originated from the magnetization dynamics in a ferrite disk, no displacement electric current should be taken into consideration and, thus, the MS description is still valid in a dielectric sample. Fig. 1 illustrates the electric-polarization dynamics in a dielectric sample placed very closely to a ferrite disk. For an observer in a laboratory frame, the magnetic and electric fields outside a MDM ferrite disk are synchronically rotating fields. The entire field structure is called a ME field [35].

It is worth noting, however, that the basic condition

$$\varepsilon_0 \frac{\partial \vec{E}}{\partial t} + \frac{\partial \vec{p}}{\partial t} << \vec{\nabla} \times \vec{m}, \tag{13}$$

necessary for the MS description, can be violated when the axes of a ferrite disk and a dielectric disk, shown in Fig. 1, are shifted or, in general, a dielectric sample is not cylindrically symmetric. In these cases the time variations of vectors $\vec{p}$ are not synchronized with the time variations of vector $\vec{m}$ and the electric polarization becomes time varying with respect to the magnetization. As a result, the electric displacement current is not negligibly small and a component of a curl magnetic field appears. This, however, does not lead to "restoration" of a basic Maxwellian field structure. We concern this problem, more in details, in the next section.

## C. Pseudoscalar helicity parameters for ME fields

In Ref. [36] it was shown that at MDM resonances, one has nonzero product $\vec{m} \cdot (\vec{\nabla} \times \vec{m})^*$. This product, characterizing the way in which the field lines of magnetization curl themselves, we call *magnetization helicity parameter*. For magnetization defined as $\vec{m} = -\vec{\tilde{\chi}} \cdot \vec{\nabla} \psi$, with an assumption that MDMs are the fields rotating with an azimuth number $\nu$ $(\psi \propto e^{-i\nu\theta})$, we obtain

$$\vec{m} \cdot (\vec{\nabla} \times \vec{m})^* = -i2C^2 \xi(z) \frac{\partial \xi(z)}{\partial z} \left( \chi \frac{\partial \tilde{\varphi}}{\partial r} + \frac{\chi_a}{r} \nu \tilde{\varphi} \right) \left( \frac{\chi}{r} \nu \tilde{\varphi} + \chi_a \frac{\partial \tilde{\varphi}}{\partial r} \right). \tag{14}$$

Here $\chi$ and $\chi_a$ are, respectively, diagonal and off-diagonal components of the magnetic susceptibility tensor $\vec{\tilde{\chi}}$ [20, 21]. We can see that the magnetization helicity parameter is purely an imaginary quantity:

$$\mathcal{V} \equiv -\operatorname{Im}\left[\vec{m}\cdot\left(\vec{\nabla}\times\vec{m}\right)^{*}\right]. \tag{15}$$

This parameter, appearing since the magnetization $\vec{m}$ in a ferrite disk has two parts: the potential and curl ones [35], can be also represented as

$$\mathcal{V} = \frac{1}{\omega\mu_0}\operatorname{Re}\left[\vec{j}^{(m)}\cdot\left(\vec{j}^{(e)}\right)^{*}\right], \tag{16}$$

where $\vec{j}^{(m)} = i\omega\mu_0\vec{m}$ and $\vec{j}^{(e)} = \vec{\nabla}\times\vec{m}$ are, respectively, the magnetic and electric current densities in a ferrite medium [1, 35]. The pseudoscalar parameter $\mathcal{V}$ gives evidence for the presence of two coupled and mutually parallel currents – the electric and magnetic ones – in a localized region of a microwave structure. The magnetization helicity parameter can be considered as a certain source which defines the helicity properties of ME fields.

An analysis of helicity properties of ME fields we should start with consideration of so-called optical chirality density. Recently, significant interest has been aroused by a rediscovered measure of helicity in optical radiation – commonly termed optical chirality density – based on the Lipkin's "zilch" for the fields [45]. The optical chirality density for propagating electromagnetic waves is defined as [2, 3]:

$$C_{opt} = \frac{\varepsilon_0}{2}\vec{E}\cdot\nabla\times\vec{E} + \frac{1}{2\mu_0}\vec{B}\cdot\nabla\times\vec{B}. \tag{17}$$

This is a time-even, parity-odd pseudoscalar parameter. Lipkin showed [45], that the chirality density is zero for a linearly polarized plane wave. However, for a circularly polarized wave, Eq. (17) gives a nonvanishing quantity. Moreover, for right- and left-circularly polarized waves one has opposite signs of parameter $C_{opt}$. It is evident that for monochromatic electromagnetic waves we have

$$\operatorname{Im}C_{opt} = \operatorname{Im}\left[\frac{\varepsilon_0}{4}\vec{E}\cdot\left(\nabla\times\vec{E}\right)^{*} + \frac{\mu_0}{4}\vec{H}\cdot\left(\nabla\times\vec{H}\right)^{*}\right] = \frac{\omega\varepsilon_0\mu_0}{2}\operatorname{Re}\left(\vec{E}^{*}\cdot\vec{H}\right) \equiv 0. \tag{18}$$

Following Ref. [46] we consider, formally, two separate terms in the left-hand side of Eq. (18). For EM fields, there are two pseudoscalar parameters

$$F^{(E)} = \operatorname{Im}\left[\frac{\varepsilon_0}{4}\vec{E}\cdot\left(\nabla\times\vec{E}\right)^{*}\right] \quad \text{and} \quad F^{(H)} = \operatorname{Im}\left[\frac{\mu_0}{4}\vec{H}\cdot\left(\nabla\times\vec{H}\right)^{*}\right]. \tag{19}$$

To satisfy Eq. (18) we may have two cases for monochromatic EM waves:

$$F^{(E)} \equiv 0 \quad \text{and} \quad F^{(H)} \equiv 0 \tag{20}$$

or

$$F^{(E)} \equiv -F^{(H)}. \tag{21}$$

Evidently, for Maxwellian fields, both Eqs. (20) and (21) are trivial.

For quasistatic ME fields, however, the situation appears quite different. Based on Eqs. (2) and (3), one obtains for the near fields outside a MDM ferrite sample [35]:

$$F^{(E)} = \frac{\varepsilon_0}{4} \text{Im}\left[ \vec{E} \cdot \left( \nabla \times \vec{E} \right)^* \right] = \frac{\omega \varepsilon_0 \mu_0}{4} \text{Re}\left( \vec{E} \cdot \vec{H}^* \right) \neq 0 \tag{22}$$

and

$$F^{(H)} \equiv 0. \tag{23}$$

We call the pseudoscalar parameter $F_{ME}^{(E)}$ the "electric" helicity of a ME field [35, 46]. This parameter gives evidence for the presence of mutually parallel components of the electric and magnetic fields in the ME-field structure. Both these, electric and magnetic, fields − the fields in the right-hand side of Eq. (22) − are potential fields. The outside electric field, however, has two components: the potential [defined by Eq. (9)] and the curl ones. The curl electric field is found from the Faraday equation

$$\vec{\nabla} \times \vec{E} = -\mu_0 \frac{\partial \vec{H}}{\partial t}, \tag{24}$$

where the magnetic field is described by Eq. (10). For a monochromatic field, the curl electric field is perpendicular to the potential magnetic field. This fields (the curl electric and potential magnetic) form power-flow vortices outside a MDM ferrite disk.

Let us suppose now that in a combined structure shown in Fig. 1, the axes of a ferrite disk and a dielectric disk are shifted or, in general, a dielectric sample is not cylindrically symmetric. The electric polarization $\vec{p}$ in this dielectric sample outside a ferrite is still induced by the electric field originated from a ferrite disk at the MDM resonance. The electric-polarization dynamics in a dielectric is strongly determined by the magnetization dynamics in a ferrite, however, the time variations of vectors $\vec{p}$ in are not synchronized with the time variations of vector $\vec{m}$ (which has the spin and orbital rotations in a ferrite). So, the condition (13) is not fulfilled.

At the MDM resonances, the entire structure (a ferrite disk and a dielectric sample) is manifested as a strongly temporally dispersive system. Assuming that a dielectric sample has sizes much less than a free-space electromagnetic wavelength, we can neglect in a dielectric a time variation of magnetic energy in comparison with a time variation of electric energy. The situation appears to be dual with respect to the above considered quasi-magnetostatic approach in a ferrite sample: we can neglect the magnetic displacement current and have a just only a potential electric field in a dielectric sample [35]. In such a quasi-electrostatic approach in a dielectric sample, the electric displacement current is not negligibly small and so the component of a curl magnetic field appears. In dielectric regions, where the condition (13) is not fulfilled, we have

$$\nabla \cdot \vec{D} = 0, \tag{25}$$

$$\vec{\nabla} \times \vec{E} = 0, \tag{26}$$

$$\vec{\nabla} \times \vec{H} = \frac{\partial \vec{D}}{\partial t}, \tag{27}$$

where $\vec{D} = \varepsilon_0 \vec{E} + \vec{p}$. In these equations, a potential electric field $\vec{E}$ is defined by Eq. (9) and a curl magnetic field is accompanied also with a potential magnetic field described by Eq. (10). Similar to the considered above a pseudoscalar parameter of the "electric" helicity, we can introduce now pseudoscalar parameter of the "magnetic" helicity, $F^{(H)}$. Inside a dielectric, we have

$$F^{(H)} = \frac{\mu_0}{4} \text{Im}\left[ \vec{H} \cdot \left( \nabla \times \vec{H} \right)^* \right] = -\frac{\omega \varepsilon_0 \mu_0}{4} \text{Re}\left( \vec{H} \cdot \vec{E}^* \right) \neq 0. \tag{28}$$

and

$$F^{(E)} \equiv 0. \tag{29}$$

The parameter $F^{(H)}$ also gives evidence for the presence of mutually parallel components of the electric and magnetic fields. Similar to the pseudoscalar parameter $F^{(E)}$ [defined by Eq. (22)], in a case of the parameter $F^{(H)}$ both the fields in the right-hand side of Eq. (28) are potential fields. It is worth noting, however, the electric and magnetic components of the ME fields (and, consequently, the entire structures of the ME fields), described by Eqs. (22), (23) from one side and Eqs. (28), (29) from the other side, are absolutely not the same. There are two different solutions: the quasi-magnetostatic and quasi-electrostatic ones. Moreover, these two different solutions are related to two different space regions.

Following Refs. [35, 46], we can introduce the normalized helicity parameters. The normalized parameter of the "electric" helicity is expressed as

$$\cos \alpha = \frac{\text{Im}\left[ \vec{E} \cdot \left( \vec{\nabla} \times \vec{E} \right)^* \right]}{\left| \vec{E} \right| \left| \nabla \times \vec{E} \right|}, \tag{30}$$

while the normalized parameter of the "magnetic" helicity is expressed as

$$\cos \beta = \frac{\text{Im}\left[ \vec{H} \cdot \left( \vec{\nabla} \times \vec{H} \right)^* \right]}{\left| \vec{H} \right| \left| \nabla \times \vec{H} \right|}. \tag{31}$$

These normalized parameters define angles between the electric- and magnetic-field components for two different structures of ME fields: the quasi-magnetostatic and quasi-electrostatic ones.

**D. ME fields and axion electrodynamics**

The ME fields, being originated from magnetization dynamics at MDM resonances, appear as the fields of axion electrodynamics [47]. In axion electrodynamics, the coupling between an axion field (which we, in general, denote $\vartheta$) and the electromagnetic field is expressed by an additional term in the ordinary Maxwell Lagrangian

$$\mathcal{L}_\vartheta = \kappa \vartheta \vec{E} \cdot \vec{B}, \tag{32}$$

where $\kappa$ is a coupling constant. This coupling results in modified electrodynamics equations with the electric charge and current densities replaced by [47 – 49]

$$\rho^{(E)} \to \rho^{(E)} + \kappa \vec{\nabla} \vartheta \cdot \vec{B}, \tag{33}$$

$$\vec{j}^{(E)} \to \vec{j}^{(E)} - \kappa \left( \frac{\partial \vartheta}{\partial t} \vec{B} + \vec{\nabla} \vartheta \times \vec{E} \right). \tag{34}$$

The axion field $\vartheta$ transforms as a pseudoscalars under space reflection *P* and it is odd under time reversal *T*. Importantly, if $\vartheta$ is a space–time constant then its contribution to the classical equations of motion vanishes.

The magnetization helicity *V*, defined by Eqs. (15), (16), is a pseudoscalar field. We reassign the magnetization helicity *V* as an axion field $\vartheta$. In the absence of free charges and conduction currents we have two equations:

$$\vec{\nabla} \cdot \vec{D} = \kappa \vec{\nabla} \vartheta \cdot \vec{B} \tag{35}$$

and

$$\vec{\nabla} \times \vec{B} - \mu_0 \varepsilon_0 \frac{\partial \vec{E}}{\partial t} = -\kappa \left( \frac{\partial \vartheta}{\partial t} \vec{B} + \vec{\nabla} \vartheta \times \vec{E} \right). \tag{36}$$

In a bulk magnetic insulator, the first term in the right-hand side of Eq. (36) can be identified as the electric-polarization current, while the second term as the magnetization current [49, 50]. The magnetization helicity, defined by Eqs. (15), (16), is a time averaged quantity. So, in Eq. (36) we have $\frac{\partial \vartheta}{\partial t} = 0$. Also, a displacement current $\mu_0 \varepsilon_0 \frac{\partial \vec{E}}{\partial t}$ is a negligibly small quantity in our consideration. Taking into account Eq. (2), the constitutive relation $\vec{B} = \mu_0 \left( \vec{H} + \vec{m} \right)$, and writing $\vartheta = \alpha \mathcal{V}$, where $\alpha$ is a dimensional coefficient, we represent Eqs. (35), (36) as

$$\rho^{(E)} = \kappa \alpha \vec{\nabla} \mathcal{V} \cdot \vec{B} \tag{37}$$

and

$$\vec{\nabla} \times \vec{m} = -\frac{\kappa \alpha}{\mu_0} \vec{\nabla} \mathcal{V} \times \vec{E}. \tag{38}$$

Eqs. (37), (38) clearly show that a MDM ferrite disk is a particle with the property of magnetoelectric polarizability. Based on these equations one can see how the distributions of electric charges and magnetization are related to the magnetization helicity factor and distributions of the electric and magnetic fields. For example, let us consider the main (the 1st radial) MDM in a ferrite disk [28]. For this mode, on a surface of a ferrite in the central region of the disk, we have for the fields are characterized by the following properties. The rotating electric-field vectors have only in-plane components [35, 43]. Also, $\vec{m}$ and $\vec{\nabla} \times \vec{m}$ are rotating in-plane vectors. $\vec{\nabla}\mathcal{V}$ has mainly the *z* component. Since vectors $\vec{m}$ and $\vec{\nabla} \times \vec{m}$ are mutually $90°$-shifted in space and time [36], we have from Eq. (38):

$$\vec{m} \propto \hat{z} \times \vec{E}, \qquad (39)$$

where $\hat{z}$ is unit vector along *z* axis. Eq. (39) shows that at the MDM resonance there is mutual correlation of the distribution of vectors $\vec{m}$ and $\vec{E}$ in a ferrite sample. It is evident that the in-plane vectors $\vec{m}$ and $\vec{E}$ are reciprocally perpendicular.

At the MDM resonances, we observe special relationships between the magnetization and electric polarization, both inside a YIG disk and in dielectrics closely abutting to the ferrite. While, in classical electrodynamics, Maxwell equations well describe the dynamic relations between electric and magnetic fields, relations between electric polarization and magnetization appear as a highly nontrivial issue. One of the basic reason is that the electrons contribute to the electric polarization and magnetic moments in completely different ways. If these two ways are specifically correlated and both the ways of contributions are long-range ordered, the ME coupling (the coupling between electric polarization and magnetization) could be enhanced. With such a ME coupling, the electric and magnetic properties should be understood and treated by a peculiar unified way.

The ME-field helicity densities, defined by Eqs. (22), (30), from the one side, and Eqs. (28), (31), from the other side, transform as pseudoscalars under space reflection *P* and are odd under time reversal *T*. It means that the ME fields are pseudoscalar axionlike fields. Whenever pseudoscalar axionlike fields, is introduced in the electromagnetic theory, the dual symmetry is spontaneously and explicitly broken. This results in non-trivial coupling between pseudoscalar quasistatic ME fields and the EM fields in microwave structures with an embedded MDM ferrite disk. A special role in this coupling plays the effect of interaction of MDM oscillations with metal surfaces.

**E. MDM oscillations in ferrite-metal structures. The ME-EM field interaction in a microwave waveguide**

In the problems, where one considers an interaction of MDM oscillations with a metal surface, the electric displacement current is also neglected. On a metal surface, however, Eq. (2) is replaced by the Ampere equation

$$\vec{\nabla}_S \times \vec{H} = \vec{j}_S. \qquad (40)$$

In the MS-wave problems, the current $\vec{j}_S$ is considered as a surface electric current which just only gives discontinuity of the tangential component of the magnetic field on a metal [51 – 53]. Even so, it is evident, however, that to define the induced electric current in the Ampere equation, the Faraday law should be used. Experimental results of magnetic-induction probing of the fields near a ferrite sample [54, 55], show that the Faraday equation yields the electric field associated with the magnetic fields of MDMs. In a view of these experiments, one can conclude that the near-field interaction of MDM resonances with external metal elements cannot be analyzed without the Faraday law.

Let us consider a flat metal screen placed in vacuum above a ferrite disk in parallel to the disk plane and at a distance much less than the disk diameter. The MDM ferrite disk is a particle with the property of ME polarizability. On a surface of a MDM ferrite disk, there are both the regions of the orbitally driven normal magnetic and normal electric fields. Moreover, the maximums (minimums) of the electric and magnetic fields normal to the disk are situated at the same places on the disk plane [36]. When a metal wall is placed closely to a ferrite-disk plane, this field structure is projected on the metal. Together with the surface electric charges induced on a metal wall by the electric field, there are also the Faraday-law eddy currents induced on a metal surface by a time derivative of a normal component of the rotating MDM magnetic field. Evidently, the induced surface electric current $\vec{j}_S$ should have two components. There are the linear currents arising from the continuity equation for surface electric charges, for which $\vec{\nabla}_S \cdot \vec{j}_S \neq 0$, and the Faraday-law eddy currents, for which $\vec{\nabla}_S \times \vec{j}_S \neq 0$. As a result, we have the sources of the linear-current and eddy-current components situated at the same place on a metal wall. Thus, the form of a surface electric current is not a closed line. It is a flat a spiral [36].

Unique properties of interaction of MDMs with a metal screen become more evident when one analyzes the angular-momentum balance conditions for MDM oscillations in a ferrite disk in a view of the ME-EM field coupling in a microwave waveguide. MDMs in a quasi-2D ferrite disk are microwave energy-eigenstate oscillations with topologically distinct structures of rotating fields and unidirectional power-flow circulations. The active power flow of the field both inside and outside a ferrite disk $\vec{P} = \frac{1}{2}\text{Re}\left(\vec{E}\times\vec{H}^*\right)$ has the vortex topology. In the MDM resonances, the orbital angular-momentum density is expressed as

$$\vec{L} = \frac{1}{2}\text{Re}\left[\vec{r}\times\left(\vec{E}\times\vec{H}^*\right)\right], \tag{41}$$

where $\vec{r}$ is a radius vector from the disk axis. For a given direction of a bias magnetic field, the power-flow circulations are the same inside a ferrite and in the vacuum near-field regions above and below the disk. Schematically, this is shown in Fig. 2.

Depending on a direction of a bias magnetic field, we can distinguish the clockwise and counterclockwise topological-phase rotation of the fields. The direction of an orbital angular-momentum of a ferrite disk

$$\vec{\mathcal{L}} = \frac{1}{2}\text{Re}\int_0^{\mathcal{R}}\left[\vec{r}\times\left(\vec{E}\times\vec{H}^*\right)\right]dr \tag{42}$$

is correlated with the direction of a bias magnetic field $\vec{H}_0$ (along +z axis or –z axis). When we consider a ferrite disk in vacuum environment, such a unidirectional power-flow circulation might seem to violate the law of conservation of an angular momentum in a mechanically stationary system. However, an angular momentum is seen to be conserved if topological properties of electromagnetic fields in the entire microwave structure are taken into account. In Ref. [36] it was shown that due to the topological action of the azimuthally unidirectional transport of energy in a MDM-resonance ferrite sample there exists the opposite topological reaction (opposite azimuthally unidirectional transport of energy) on a metal screen placed near this sample. This effect is called topological Lenz's effect. In a microwave structure with an embedded ferrite disk, an orbital angular momentum, related to the power-flow circulation, must be conserved in the process. Thus, if power-flow circulation is pushed in one direction in a ferrite disk, then the power-flow circulation on metal walls to be pushed in the other direction by the same torque at the same time. Fig. 3 illustrates the orbital angular-momentum balance conditions at a given direction of a bias magnetic field.

In Ref. [36], an analysis of conservation of an angular momentum was made in microwave waveguiding structures with the metal walls situated very close to the ferrite-disk surfaces. A thin ferrite disk is placed inside a rectangular waveguide symmetrically to its walls so that the disk axis is perpendicular to a wall, as it is shown in Fig. 4. Vacuum gaps between the metal and ferrite are much less than a diameter of a MDM ferrite disk and thus the entire microwave structure (a ferrite disk and a waveguide) can be considered as a quasi-2D structure. In this structure, a role of a linear EM-wave momentum is negligibly small. So, the orbital angular-momentum balance does not depend, actually, on the direction of the electromagnetic wave propagation in a waveguide. With taking into account that MDMs in a quasi-2D ferrite disk are microwave energy-eigenstate oscillations, we have the angular momentum quantization as well. The observed properties of interaction between a MDM ferrite disk and a metal screen rely on ME virtual photons to act as the mediator. There are two near-field elements of this interaction: (*a*) the static electric force and (*b*) the electromagnetic induction.

However, the fields in a ferrite disk rotate at microwave frequencies and situation becomes more complicated when the vacuum-region scale is about the disk diameter or more and the finite speed of wave propagation in vacuum − the retardation effects − should be taken into consideration. This means that for a brief period, the total angular momentum of the two topological charges (one in a ferrite, another on a metal) is not conserved, implying that the difference should be accounted for by an angular momentum in the fields in the vacuum space in a waveguide. It means that the magnetization dynamics have an impact on the phenomena connected with fluctuation energy in vacuum. As the rotational symmetry is broken in this case, the Casimir torque [56 – 60] arises because the Casimir energy now depends on the angle between the directions of the magnetization vectors in a ferrite and electric-current vectors on a metal wall. A vacuum-induced Casimir torque allows for torque transmission between the ferrite disk and metal wall avoiding any direct contact between them.

An experimental proof of the predicted above Casimir-torque effect in microwaves is an important problem. However, even in optics, such an experiment appears as an open question. While the Casimir force has been measured extensively, the Casimir torque has not been observed experimentally though it was predicted over forty years ago. Some ideas proposed recently to detect the Casimir torque with an optics (see, e. g. [61] and references therein) can be useful for the proposal of analogous experimental methods in microwaves.

### III. EVIDENCE FOR QUANTIZED STATES OF THE ME FIELDS ORIGINATED FROM MDM OSCILLATIONS

The eigenvalues of MDM resonances are defined by discrete quantities of a diagonal component of the permeability tensor [26 – 28, 34]. For a ferrite sample, magnetized at saturation magnetization $M_0$, the diagonal component of the tensor (5) is found as [20]

$$\mu = 1 + \frac{\gamma^2 \mu_0^2 M_0 H_i}{\gamma^2 \mu_0^2 H_i^2 - \omega^2}, \tag{43}$$

where $\gamma$ is the gyromagnetic ratio and $H_i$ is a DC internal magnetic field. In neglect of material anisotropy, the internal magnetic field is calculated as

$$\vec{H}_i = \vec{H}_0 - \vec{H}_d, \tag{44}$$

where $H_0$ is a bias magnetic field and $H_d$ is a demagnetization field [20]. An analysis [26 – 28, 34] shows that real energy-eigenstate solutions for MS-potential wave functions in a quasi-2D ferrite disk are obtained when $\mu < 0$. This corresponds to the frequency range [20]:

$$\omega_H < \omega < \omega_\perp, \qquad (45)$$

where $\omega_H = \mu_0 \gamma H_i$, $\omega_\perp = \sqrt{\omega_H (\omega_H + \omega_M)}$. Here $\omega_M = \mu_0 \gamma M_0$. The magnetic-field range for negative quantities $\mu$ are defined as [20]

$$H_i^\diamond < H_i < \omega / \mu_0 \gamma, \qquad (46)$$

where $H_i^\diamond \equiv \sqrt{\left(\dfrac{\omega}{\mu_0 \gamma}\right)^2 + \left(\dfrac{M_0}{2}\right)^2} - \dfrac{M_0}{2}$.

Sharp multiresonance oscillations, observed experimentally in microwave structures with an embedded quasi-2D ferrite disk [29 – 32], are related to magnetization dynamics in the sample. This dynamics have an impact on the phenomena connected with the quantized energy fluctuation. For given sizes of a disk and a given quantity of saturation magnetization of ferrite material $M_0$, there are two different mechanisms of the MDM energy quantization: (*i*) by a signal frequency $\omega$ at a constant bias magnetic field $H_0$ and (*ii*) by a bias magnetic field $H_0$ at a constant signal frequency $\omega$. Fig. 5 shows the FMR diagonal component of the permeability tensor $\mu$ versus frequency $\omega$ at a constant internal magnetic field $H_i$. The FMR diagonal component of the permeability tensor versus an internal magnetic field at a constant frequency is shown in Fig. 6. For both these cases, the discrete quantities of $\mu$, in the region where $\mu < 0$, are shown schematically for the first four MDM resonances.

Fig. 7 illustrates correlation between the two mechanisms of the MDM energy quantization. For a certain frequency $f'$, the energy quantization is observed at specific bias-field quantities: $\left(H_0^{(1)}\right)'$, $\left(H_0^{(1)}\right)'$, $\left(H_0^{(1)}\right)'$, …. Evidently, one has to use a statistical description of the spectral response functions of the system with respect to two external parameters – a bias magnetic field $H_0$ and a signal frequency $\omega$ – and analyze the correlations between the spectral response functions at different values of these external parameters. It means that, in neglect of losses, there should exist a certain *uncertainty limit* stating that

$$\Delta f \Delta H_0 \geq \text{uncertainty limit}. \qquad (47)$$

This uncertainty limit is a constant which depends on the disk size parameters and the ferrite material property (such as saturation magnetization). Beyond the frames of the uncertainty limit (47) one has continuum of energy. The fact that there are different mechanisms of quantization allows to conclude that for MDM oscillations in a quasi-2D ferrite disk both discrete energy eigenstate and a continuum of energy can exist. In quantum mechanics, the uncertainty principle says that the values of a pair of canonically conjugate observables cannot both be precisely determined in any quantum state. In a formal harmonic analysis in classical physics, the uncertainty principle can be summed up as follows: A nonzero function and its Fourier transform cannot be sharply localized. This principle states also that there exist limitations in performing measurements on a system without disturbing it. Basically, formulation of the main statement of the MDM-oscillation theory is impossible without using a classical microwave structure. If a MDM particle is under interaction with a "classical electrodynamics" object, the states of this classical object change. The character and value of these changes depend on the MDM quantized states and so can serve as its qualitative characteristics. The microwave measurement reflects interaction between a microwave structure and a MDM particle. It is worth noting that for different types of subwavelength particles, the uncertainty principle may acquire different forms. An interesting variant of Heisenberg's uncertainty principle was shown recently in subwavelength optics [62]. Being applied to the optical field, this principle says that we can only measure the electric or the magnetic field with accuracy when the volume in which they are contained is significantly smaller than the wavelength of light in all three spatial dimensions. As volumes smaller than the wavelength are probed, measurements of optical energy become uncertain, highlighting the difficulty with performing measurements in this regime. From this statement, we can see, once again, that a MDM particle, distinguished by strong ME properties in subwavelength microwaves, is beyond the regular EM-field description.

The fact that magnetization dynamics in a quasi-2D ferrite disk have an impact on the energy quantization of the fields in a microwave cavity, was confirmed experimentally in Ref. [32]. In this work, the MDM-originated quantized states of the cavity fields were investigated with variation of a bias magnetic field at a constant operating frequency, which is a resonant frequency of the cavity. The observed discrete variation of the cavity impedances are related to discrete states of the cavity fields. Since the effect was obtained at a certain resonant frequency, the shown resonances are not the modes related to quantization of the photon wave vector in a cavity. It is evident that these resonances should be caused by the quantized variation of energy of a ferrite disk, appearing due to variation of energy of an external source– a bias magnetic field. In Ref. [32], the observed effect of energy quantization of the fields in a microwave structure in relation to quantization of magnetic energy in a ferrite disk is analyzed qualitatively as follows. At the regions of a bias magnetic field, designated in Fig. 8 as *A, a, b, c, d,* …, we do not have MDM resonances. In these regions, a ferrite disk is "seen" by electromagnetic waves, as a very small obstacle which, practically, does not perturb a microwave cavity. In this case, the cavity (with an embedded ferrite disk) has good impedance matching with an external waveguiding structure and a microwave energy accumulated in a cavity is at a certain maximal level. At the MDM resonances (the states of a bias magnetic fields designated in Fig. 8 by numbers 1, 2, 3, …), the reflection coefficient sharply increases. The input impedances are real, but the cavity is strongly mismatched with an external waveguiding structure. It means that at the MDM resonances, the cavity accepts smaller energy from an external microwave source. In these states of a bias magnetic field, the microwave energy accumulated in a cavity sharply decreases, compared to its maximal level in the *A, a, b, c, d,* … Since the only external parameter, which varies in this experiment, is a bias magnetic field, such a sharp ejection of the microwave energy accumulated in a cavity to an external waveguiding structure should be related to emission of discrete portions of energy from a ferrite disk. It means that at the MDM resonances, we should observe strong and sharp reduction of magnetic energy of a ferrite sample.

The question on a proper explanation of the experimentally observed quantized states of the fields in a microwave structure with an embedded ferrite disk remains still open. In Refs. [29, 30, 63], it was stated that the main reason of appearing the multiresonance oscillations is non-homogeneity of an internal DC magnetic field in a ferrite-disk sample. To answer the question whether non-homogeneity of an internal magnetic field can really lead to quantization of magnetic energy of a ferrite disk, we analyze briefly the model suggested in Ref. [30] and modified in Ref. [64].

In a supposition that saturation magnetization $M_0$ is uniform everywhere inside the sample, the demagnetization magnetic field in a quasi-2D ferrite disk varies only in the radial direction and is defined as [30]

$$\vec{H}_d(r) = -\vec{M}_0 I(r), \qquad (48)$$

where $I(r)$ is a dimensionless scalar quantity and $0 \leq r \leq \mathcal{R}$. In Ref. [30], the multiresonance absorption peaks are interpreted to be caused by MS waves propagating radially across the disk with the internal magnetic field dependent on a radial coordinate in the plane of a YIG film and the mode numbers are determined based on the well-known Bohr–Sommerfeld quantization rule. This model is illustrated in Figs. 9 and 10. Fig. 9 shows schematically the Bohr–Sommerfeld quantization rule for a certain MDM. A qualitative picture of the levels of an internal magnetic field for standing waves corresponding to some MDMs is shown in Fig. 10. With increasing the mode number, the effective diameter $2\mathcal{R}_{eff}^{(n)}$ of the disk increases as well. The ''in-plane'' MS-potential-function distribution is supposed to be azimuthally nondependent.

However, the Bohr–Sommerfeld quantization used in Ref. [30] does not clarify the problem of the observed quantization of magnetic energy of a ferrite-disk sample. This quantization should be observed as a Zeeman splitting of the energy levels created by a DC internal magnetic field. In an assumption that saturation magnetization $M_0$ is uniform inside the sample and is the same for every mode, one cannot suggest any mechanism of Zeeman splittings at the MDM resonances. When a ferrite disk is placed in a homogeneous external (bias) magnetic field $H_0$, magnetic energy of the entire sample is varied monotonically with variation of $H_0$, even if the MS-wave numbers are quantized and the effective diameter $2\mathcal{R}_{eff}^{(n)}$ is quantized as well.

Magnetic energy of a sample in an external (bias) magnetic field is determined by the demagnetization field. The demagnetization field is the magnetic field generated by the magnetization in a magnet and the demagnetization factor determines how a magnetic sample responds to an external (bias) magnetic field. It is evident that for the MDM spectra, obtained at variation of a bias magnetic field $H_0$ and at a constant signal frequency $\omega$, a discrete reduction of magnetic energy of a ferrite disk at the MDM resonance should occur because of quantization of the demagnetization field. Contrary to Ref, [30], will consider of quantization of the demagnetization field due to quantization of DC magnetization. Following the technique described in Ref. [64], we can put aside the question on a role of nonhomogeniety of an internal magnetic field in a ferrite disk. With averaging of the parameter $I(r)$ on the region of the actual diameter of the disk, $2\mathcal{R}$, one can write [64]

$$(H_i)_{average} = H_0 - (H_d)_{average}, \qquad (49)$$

where

$$(H_d)_{average} = -M_0 I_{average}. \qquad (50)$$

Here $0 < I_{average} < 1$. The quantity of $I_{average}$ depends on the disk geometry. With an approximation of a homogeneous (averaged on the disk-diameter region) internal magnetic field, one can obtain the diagonal permeability-tensor component $\mu$ from Eq. (43). The spectral problem is analyzed with $\psi$-function distributions in a form of Eq. (6). The eigenvalues of MDM resonances are defined by discrete quantities of the diagonal permeability-tensor component $\mu$. In this case, one can classify MDMs as the thickness, radial, and azimuthal modes. The spectral problem for these modes gives evidence for energy eigenstate oscillations [26, 28]. The technique used in Ref. [64] gives a good agreement with experimental results of the resonance mode position in the spectra.

For simplicity of a further analysis, we will suppose that $I_{average} = 1$. Assuming that quantization of magnetic energy of a ferrite-disk sample is due to quantization of the DC magnetization, the demagnetization magnetic field for a certain MDM with number $n$ ($n$ = 1, 2, 3, …), is found as

$$H_d^{(n)} = -\left(M_0^{(n)}\right)_{eff},\tag{51}$$

where $\left(M_0^{(n)}\right)_{eff} \leq M_0$ is a quantized DC magnetization. We can write

$$\left(M_0^{(n)}\right)_{eff} \equiv M_0 K^{(n)},\tag{52}$$

where the mode coefficient is a dimensionless quantity: $0 \leq K^{(n)} \leq 1$. The number $n = 1$ corresponds to the main MDM [28].

For mode $n$, the magnetic energy of a sample of volume $V$, placed in a magnetic field $H_0$, is calculated as

$$-W^{(n)} = \frac{1}{2}\int_V \vec{H}\cdot\vec{M}\,dV = \frac{1}{2}H_0\left(M_0^{(n)}\right)_{eff} V.\tag{53}$$

Discreteness of magnetic energy in a ferrite disk is due to discrete reduction:

$$\Delta W^{(n)} = -\frac{1}{2}\int_V \vec{H}\cdot\Delta\left(\vec{M}^{(n)}\right)_{eff} dV = -\frac{1}{2}H_0\left(1-K^{(n)}\right)M_0 V.\tag{54}$$

The energy $\Delta W^{(n)}$ is the microwave energy extracted from the magnetic energy of a ferrite disk at the $n$-th MDM resonance. The effect is illustrated in Fig. 11. Based on this model, we can explain qualitatively how the multiresonance states in a microwave cavity, experimentally observed in Ref. [32], are related to quantized variation of energy of a ferrite disk, appearing due to an external source – a bias magnetic field. When accepting this model, we can say that we have a quantum effect of electromagnetically generated demagnetization of a sample.

What is the physics of quantization of a DC magnetization in a ferrite disk? Unidirectionally rotating fields with the spin and orbital angular momenta, observed at MDM resonances, are related to precessing magnetic dipoles in a ferrite and to a double-valued-function magnetic current on a lateral surface of a ferrite disk [34, 35]. Circulation of the chiral surface magnetic current results in appearance of a DC gauge electric field [34] and thus appearance of DC electric charges on the ferrite-disk planes. When a dielectric-disk sample is situated in a vacuum near a ferrite disk, it is subjected with an *induced electric gyrotropy and orbitally driven electric polarization* [35]. Because of the electric field originated from a ferrite disk, every separate electric dipole in a dielectric disk precesses around its own axis and for all the precessing dipoles, there is an orbital phase running (see Fig. 1). Due to the orbital angular momentum of MDM oscillations, a torque exerting on the electric polarization in a dielectric sample should be equal to a reaction torque exerting on the magnetization in a ferrite disk. Because of this reaction torque, the precessing magnetic moment density of the ferromagnet will be under additional mechanical rotation at a certain frequency. At dielectric loadings, the magnetization motion in a ferrite disk is characterized by an effective magnetic field [35, 65]. This is a DC gauge (topological) magnetic field caused by precessing and orbitally rotating electric dipoles in a dielectric sample. As a result, an *external* effective magnetic field becomes less than a bias magnetic field $H_0$. So, at a dielectric loading of a ferrite sample, the Larmor frequency $\omega_H$ should be lower than such a frequency for an unloaded ferrite disk. In other words, one can say that due to a loading by an external dielectric sample, the effective DC magnetic charges appear on the ferrite-disk planes.

At the same time, it is worth noting that a ferrite disk is made of a magnetic *dielectric* – yttrium iron garnet (YIG) – which has sufficiently high permittivity, $\varepsilon_r = 15$. Inside the YIG disk, we also have the torque exerting on the electric polarization due to the magnetization dynamics of MDM oscillations. Because of the effective magnetic charges on a ferrite-disk planes (caused now by the induced electric gyrotropy and orbitally driven electric polarization inside a ferrite), the demagnetizing magnetic field is reduced. It means that the DC magnetization of a ferrite disk is reduced as well. We have the frequency $\left(\omega_M^{(n)}\right)_{eff} = \gamma\mu_0 \left(M_0^{(n)}\right)_{eff}$, which is less than such a frequency $\omega_M = \gamma\mu_0 M_0$ in an unbounded magnetically saturated ferrite. In connection with the effect discussed above, it is relevant to refer here to some recent studies in optics. In Ref. [66], light-induced magnetization using nanodots and chiral molecules was experimentally studied. It was shown that a torque transferred through the chiral layer to a ferromagnetic layer, can create local perpendicular magnetization. The experiment is based on the chiral-induced spin-selectivity effect described in Refs. [67 – 70]. An important conclusion arises also from the above consideration. When at a constant frequency of a microwave signal we vary a bias magnetic field, at MDM resonances we can observe a DC magnetoelectric effect. We have quantized DC electric and magnetic charges on the ferrite-disk planes.

Experiments in Ref. [32] shows that for the quantized states of microwave energy in a cavity and magnetic energy in a ferrite disk, the cavity input impedances on the complex-reflection-coefficient plane (the Smith chart) [71] are real numbers (see Fig. 12). We have a two-state system. As the energy swept through an individual resonance, one observes evolution of the phase – the phase lapses. From the Smith chart one can also see that the phase jump of $\pi$ is observed each time a resonant condition is achieved. In a 2D parametric space – the impedance space on a Smith chart – there is clockwise or counter clockwise circulation for the MDM states. Direction of the circulation should be correlated with the direction of a bias magnetic field $H_0$. In Fig. 12 we, contingently, showed the clockwise circulation.

At the first glance, the analyzed above processes in a structure shown in Fig. 8 can be described based on a simple scheme shown in Fig. 13. At a given frequency $\omega$, determined by a RF source, a microwave waveguide is presented by a characteristic impedance $Z_0$. The waveguide is loaded by an impedance $Z_L$, which depends on an external parameter – a bias magnetic field $H_0$. Such a simple scheme, however, leads us to an evident contradiction. With acceptance of the fact that the microwave energy extracted from the magnetic energy of a ferrite disk at the MDM resonance is a quantized quantity, we cannot assume, at the same time, that the frequency of a microwave photon propagating in a waveguide is a constant quantity.

## IV.   MDM RESONANCES AND BOUND STATES IN THE MICROWAVE CONTINUUM

When analyzing the scattering of EM waves by MDM disks in microwave waveguides and energy quantization of the field in a microwave cavity, it is relevant also to dwell on some basic problems of magnon-photon interaction and bound states in the microwave continuum. We are witnesses that long-standing research in coupling between electrodynamics and magnetization dynamics noticeably reappear in recent studies of magnon-photon interaction. In a series of works [72 – 76], it was shown that magnetostatic modes in a small YIG *sphere* can coherently interact with photons in a microwave cavity. In a small ferromagnetic particle, the exchange interaction can cause a very large number of spins to lock together into one macrospin with a corresponding increase in oscillator strength. This results in strong enhancement of spin-photon coupling relative to paramagnetic spin systems [72]. The total Hamiltonian of the system incorporates the magnetic $\vec{H}$ and electric $\vec{E}$ fields of the cavity and the $\vec{M}$ magnetization of the ferromagnetic particle [1]

$$\mathcal{H} = \frac{1}{2} \int \left[ \mu_0 \left| \vec{H} \right|^2 + \varepsilon_0 \left| \vec{E} \right|^2 + \mu_0 \left( \vec{H} \cdot \vec{M} \right) \right] d^3 r . \tag{55}$$

The spatially uniform mode of the magnetization dynamics is called the Kittel mode. The Kittel and cavity modes form magnon-polariton modes, i.e., hybridized modes between the collective spin excitation and the cavity excitation. In the theory, the interaction between the microwave photon and magnon is described by the Hamiltonian with a rotating-wave approximation (RWA) [72, 73]. In a structure of a microwave cavity with a YIG sphere inside, the avoided crossing in the microwave reflection spectra (obtained with respect to the bias magnetic field) verifies the strong coupling between the microwave photon and the magnon. The Zeeman energy is defined by a coherent state of the macrospin/photon system when a magnetic dipole is in its antiparallel orientation to the cavity magnetic field. The coherent energy exchange occurs back and forth between photon and a macrospin states. It is pointed out that because of non-uniformity of the RF magnetic field in sufficiently big YIG spheres, high-order magnon modes – the MDMs – can be excited. For such MDMs, the avoided crossing in the reflection spectra, obtained with respect to the bias magnetic field, is observed as well [73 – 76]. Importantly, for these modes, characterizing by non-uniform magnetization dynamics, the model of coherent states of the macrospin/photon system discussed above, is not applicable.

   The MDMs in YIG spheres were observed long ago [77]. There are so-called Walker modes [37]. While for a ferrite sphere one sees a few broad MDM absorption peaks, the MDMs in a quasi-2D ferrite disk are presented with the spectra of multiresonance sharp peaks. There are very rich spectra of both types, Fano and Lortenzian, of the peaks [29 – 32]. The ferrite disk is an open hi-Q resonator embedded in a microwave waveguide or microwave cavity. In such a structure, sharp MDM resonances can appear as bound states in a microwave-field continuum.

   Bound states in the continuum (BICs), also known as embedded trapped modes, are localized solutions which correspond to discrete eigenvalues coexisting with extended modes of a continuous spectrum. The BICs are solutions having an infinitely long lifetime. Recent developments show that in a large variety of electromagnetic structures there can be different mechanisms that lead to BICs [78]. One of the main reasons for appearance of the MDM BICs is a symmetry mismatch. Modes of different symmetry classes (such as reflection or rotation) are completely decoupled. MDM oscillations do not exhibit a rotational symmetry, while a regular waveguide structure is rotationally symmetric. With such a condition, the MDM bound states observed in microwave structures can be classified as the symmetry-protected BICs.

The MDM bound states are embedded in the microwave continuum but not coupled to it. Certainly, if a bound state of one symmetry class is embedded in the continuous spectrum of another symmetry class their coupling is forbidden. Moreover, the ME fields, originated from a MDM ferrite disk, and the EM fields in a microwave structure are described by different types of equations. At any stable state, MDMs cannot radiate because there is no way to assign a far-field EM-wave polarization that is consistent with vortex ME fields near a MDM ferrite disk. In a short-range interaction, an important aspect concerns the topological nature of the MDM BICs. These topological properties can be understood through eigen power-flow vortices with corresponding topological charges [34 – 36]. Quantized topological charges cannot suddenly disappear. They are protected by special boundary conditions in a quasi-2D ferrite disk. The MDM BICs cannot be removed unless MDM topological charges are cancelled with another structure carrying the opposite topological charges. Such opposite topological charges appear on metal walls of a microwave waveguide. The probing of these BICs is due to topological-phase properties of MDMs resulting in appearance of spiral electric currents induced on metal parts of a microwave structure [36]. The coupling is possible via the continuum of decay channels at the condition that phases of the MDM bound states are strongly determined phases.

The region where a MDM ferrite disk is situated in a microwave waveguide is a contact region. In this region, the conditions of tunneling and pairwise coalescence of waveguide complex modes can be fulfilled. A model of interaction of MDM resonances with the microwave-waveguide continuum is shown schematically in Fig. 14. In the system, the whole wave function can be divided into an internal, localized MDM system $Q$ and an external part of environmental microwave states $P$ [79 – 81]. The regions in a space between a ferrite disk and waveguide walls are contact regions. Surface electric currents on waveguide walls are edge states. For a whole spectrum of waveguide modes, there should be a continuum of such edge states. A ferrite disk is considered as a defect interacting with these edge states. We suppose that the region where the defect is localized, is small compared with the characteristic length scale of waveguide modes. Interaction of MDM resonances with the microwave-waveguide continuum is realized by two ways (channels): (*i*) interaction of MDMs with metal walls and (*ii*) coupling of surface helical bound states (induced on the walls at MDM resonances [36]) with a continuum of waveguide edge states on metal walls. In a structure of a thin rectangular waveguide with an embedded quasi-2D ferrite disk [36], the contact regions can be considered as two vacuum cylinders with the same diameter as a ferrite disk. This model is inapplicable in a case of a "thick" microwave waveguide [82, 83]. However, use of two dielectric cylinders with a high dielectric constant allows considering these cylinders as the contact regions in a "thick" microwave waveguide. Such structures are shown in Ref. [83]. In the contact regions, above and below a ferrite disk, we have helical-mode tunneling. There is an evidence for a torsion structure of the fields in the contact regions [36, 83]. Rotations of the power-flow vortices along an axis of contact regions are at different directions. The model is shown in Fig. 15.

The appearance of BICs is directly related to the phenomenon of an avoided level crossing of neighbored resonance states. BICs can occur due to the direct and via-the-continuum interaction between quasistationary states and can be viewed as resonances with practically infinite lifetimes. In a system of one open channel and two discrete resonances, these resonances can be coupled via the continuum. There is a short-range interaction with a purely imaginary coupling term. If two resonances pass each other as a function of a certain continuous parameter, one of the resonance states can acquire zero resonance width. This resonance state becomes a BIC. Whether or not two resonances interfere is not directly related to whether or not they overlap. Such a mechanism of cutting down of a discrete eigenstate from any connection with a continuum was described initially in Ref. [84] in a framework of a two-level model. In electromagnetic systems, the phenomenon of avoided crossing of narrow resonance states under the influence of their coupling to the continuum is considered in many recent publications (see Ref. [78] and references therein). The BICs remain perfectly confined without any radiation. For an open resonance structure, we have non-Hermitian effective Hamiltonian with complex eigenvalues. The BIC can be found by the condition that at least one of the complex eigenvalues becomes real.

From a number of characteristic features inherent in the BICs, one should distinguish such a property as a Fano resonance collapse. For the MDM resonances the effect of Fano resonance collapse was clearly demonstrated in Ref. [85]. Thanks to the tenability of the ferrite-disk resonator by an external parameter (the bias magnetic field, for example) the MDM oscillations can become close interacting modes. Fig. 16 shows how the physical parameter – a bias magnetic field – tunes the shape of the MDM resonance. The structure used in Ref. [85] is a microwave cavity with an embedded thin-film ferrite disk. It is seen that as we approach the top of the cavity resonance curve, the minimum of the microwave transmission approaches the maximum of the transmission. At the top of the cavity resonance curve, these levels (minimum and maximum transmission) are in contact. The Fano line shape is completely damped and one observes a single Lorentzian peak. The scattering cross section corresponds to a pure dark mode. What are the neighbored resonance states in this MDM oscillation? There are the dipole (dark) and quadrupole (bright) resonances [86]. Such resonances are shown in Fig. 17. These resonances, being energy degenerated, appear due to orbital angular momentums originated from edge magnetic currents on a lateral surface of a ferrite disk [34].

When the contact regions – the space above and below a ferrite disk – are dielectric cylinders with a high dielectric constant [83], the coupling of two resonance states is via the continuum of decay channels in a dielectric region. With variation of a permittivity of dielectric cylinders $\varepsilon_r$ loading a ferrite disk, discrete eigenstates can be cut down from any connection with a continuum at a certain threshold parameter $\varepsilon_r$. This effect of the Fano resonance collapse in such a structure is shown in Fig. 18. One can see that at $\varepsilon_r = 38$, the Fano line shapes are completely damped for the 1st and 2nd MDMs and single Lorentzian peak appear. The scattering cross section corresponds to pure bright modes.

One of the features attributed to the BICs is also strong resonance field enhancement. In the case of MDM oscillations in a microwave-field continuum, such a strong field enhancement is shown in numerous numerical studies [32, 35, 82, 86, 87]. For example, Fig. 19 shows passing the front of the electromagnetic wave, when the frequency is (*a*) far from the frequency of the MDM resonance and (*b*) at the frequency of the MDM resonance in the disk. This is an evidence for strong resonance field enhancement at the MDM resonance.

Recently, the MDM BIC phenomena, found further development in a novel technique based on the combination of the microwave perturbation method and the Fano resonance effects observed in microwave structures with embedded small ferrite disks [85, 88]. When the frequency of the MDM resonance is not equal to the cavity resonance frequency, one gets Fano transmission intensity. If the MDM resonance frequency is tuned to the cavity resonance frequency, by a bias magnetic field, one observes a Lorentzian line shape. The effect of Fano resonance collapse has no relations to the quality factor of a microwave cavity. Use of an extremely narrow Lorentzian peak allows exact probing of the resonant frequency of a cavity loaded by a high lossy material sample. With variation of a bias magnetic field, one can see different frequencies of Lorentzian peaks for different kinds of material samples. This gives a picture of precise spectroscopic characterization of high absorption matter in microwaves, including biological liquids. Importantly, there is no influence of the dissipation effects in the microwave cavity on the quality of the MDM resonances. The poles in the transmission amplitudes are connected with the bound states and their lifetimes.

## V. *PT*-SYMMETRY OF MDMS IN A FERRITE DISK

A MDM ferrite disk is an open resonant system. In the description of such a system, a non-Hermitian effective Hamiltonian can be derived from the Hermitian Hamiltonian including the environment. Because of strong spin–orbit interaction in the MDM magnetization dynamics in ferrite disks, the fields of these oscillations have helical structures. The right-left asymmetry of MDM rotating fields is related to helical resonances: The phase variations for resonant $\psi$ functions are both in azimuth $\theta$ and in axial $z$ directions. This shows that proper solutions of a spectral problem for MDMs should be obtained in a helical-coordinate system. The helices are topologically nontrivial structures, and the phase relationships for waves propagating in such structures could be very special. Unlike the Cartesian- or cylindrical-coordinate systems, the helical-coordinate system is not orthogonal, and separating the right-handed and left-handed solutions is admitted. In a helical coordinate system, the solution for MS-potential wave function for MDM oscillations in an open ferrite-disk resonator have four components. In a single-column matrix, this components are presented as

$$\psi = \begin{pmatrix} \psi^{(1)} \\ \psi^{(2)} \\ \psi^{(3)} \\ \psi^{(4)} \end{pmatrix}, \tag{56}$$

where the components of the matrix are distinguished as forward right-hand-helix wave, backward right-hand-helix wave, forward left-hand-helix wave, and backward left-hand-helix wave [89].

For helical modes, there are no properties of parity (*P*) and time-reversal (*T*) invariance— the *PT* invariance. The *PT*-symmetry breaking does not guarantee real-eigenvalue spectra. It was shown, however, that by virtue of the quasi-2D of the problem for a thin-film ferrite disk, one can reduce solutions from helical to cylindrical coordinates with a proper separation of variables. As shown [89], for the double-helix resonance, one can introduce the notion of an effective membrane function $\tilde{\varphi}$. This allows reducing the problem of parity transformation to a one-dimensional reflection in space and describe the boundary-value-problem solution for the MS-potential wave function by Eq. (8). In such a case, the integrable solutions for MDMs in a cylindrical-coordinate system can be considered as *PT* invariant.

At MDM resonances, we have BICs with two degenerated real-eigenvalue resonances and an imaginary-term coupling. The Fano-resonance collapse, observed at variation of a quantity of a bias magnetic field, corresponds to the BIC state. BICs are considered as resonances with zero leakage and zero linewidth (in other words, a resonance with an infinite quality factor). In our case of MDM oscillations, BICs appear when a microwave structure is *PT*–symmetrical. A *PT*-symmetric Hamiltonian is in general not Hermitian, but if the corresponding eigenstates are also *PT*-symmetric, then the eigenvalues are real and eigenstates may be complete [90, 91]. In *PT*-symmetry MDM structures, the time direction is defined by the direction of the spin and orbital rotations of the fields in a ferrite disk [35, 43]. Depending on a direction of a bias magnetic field, we can distinguish the clockwise and counterclockwise topological-phase rotation of the fields. The direction of an orbital angular-momentum expressed by Eq. (42) is correlated with the direction of a bias magnetic field $\vec{H}_0$ (along +*z* axis or –*z* axis), determines the time direction in our "topological clock". Both directions of rotation are equivalent and the biorthogonality relations for MDMs can be used. Importantly, the MDM resonance in a *PT* symmetrical microwave structure can reveal the properties analogous to the effect of spectral singularities of complex scattering potentials in *PT*-symmetric gain/loss structures [90]. The wave incident on a MDM ferrite disk induces outgoing (transmitted and reflected) waves of considerably enhanced amplitude. The disk then uses a part of the energy of the magnetized ferrite sample to produce and emit a more intensive electromagnetic wave. This spectral singularity-related resonance effect relies on the existence of a localized region with a *PT*-symmetric complex scattering potential. Because this is a characteristic property of resonance states, spectral singularities correspond to the resonance states having a real energy.

In Refs. [34, 35], we analyzed two spectral models for the MDM oscillations in a quasi-2D ferrite disk. There are models based on the so-called $\hat{G}$ and $\hat{L}$ differential operators – the *G*- and *L*-mode spectral solutions, respectively. For *G* modes one has the Hermitian Hamiltonian for MS-potential wave functions. These modes are related to the discrete energy states of MDMs. The *G*-mode orthogonality condition is expressed as [34]

$$(E_p - E_q) \int_S \tilde{\eta}_p \tilde{\eta}_q^* dS = 0, \tag{57}$$

where the sign * means complex conjugation, $\tilde{\eta}$ is the *G*-mode MS-potential membrane function, *E* is normalized energy of the mode, and *S* is an area of the disk plane. In a case of the *L* modes, we have a complex Hamiltonian for MS-potential wave functions. For eigenfunctions associated with such a complex Hamiltonian, there is a nonzero Berry potential (meaning the presence of geometric phases) [34, 35]. The main difference between the *G*- and *L*-mode solutions becomes evident when one considers the boundary conditions on a lateral surface of a ferrite disk. In solving the energy-eigenstate spectral problem for the *G*-mode states, the boundary conditions on a lateral surface of a ferrite disk of radius $\mathcal{R}$ is expressed as

$$\mu \left( \frac{\partial \tilde{\eta}}{\partial r} \right)_{r=\mathcal{R}^-} - \left( \frac{\partial \tilde{\eta}}{\partial r} \right)_{r=\mathcal{R}^+} = 0. \tag{58}$$

This boundary condition, however, manifests itself in contradictions with the electromagnetic boundary condition for a radial component of magnetic flux density $\vec{B}$ on a lateral surface of a ferrite-disk resonator. Such a boundary condition, used in solving the resonant spectral problem for the *L*-mode states, is written as $\mu (H_r)_{r=\mathcal{R}^-} - (H_r)_{r=\mathcal{R}^+} = -i\mu_a (H_\theta)_{r=\mathcal{R}^-}$, where $H_r$ and $H_\theta$ are, respectively, the radial and azimuth components of a magnetic field on a border circle. In the MS description, for the *L*-mode membrane function $\tilde{\varphi}$, this boundary condition on a lateral surface is described by Eq. (7). While the *G* modes are single-valued-function solutions, the *L* modes are double-valued-function solutions. It follows from the fact that at a given direction of a bias magnetic field, the two (clockwise and counterclockwise) types of resonant solutions for *L*-mode states may exist.

When a ferrite disk is placed in a rectangular waveguide, so that the disk plane is in parallel to the wide waveguide walls and the disk position is symmetrical to these walls, MDM oscillations are *PT* symmetric. It follows from the property of *PT*-invariance of the *L*-mode membrane function $\tilde{\varphi}$. For a ferrite-disk axis directed along a coordinate axis *y*, on a lateral boundary of a ferrite disk we have the following condition [86]

$$\mathcal{PT} \left[ \tilde{\varphi}_{r=\mathcal{R}}(z) \right] = \tilde{\varphi}_{r=\mathcal{R}}^*(-z) = \tilde{\varphi}_{r=\mathcal{R}}(z). \tag{59}$$

The biorthogonality relation for the two $L$ modes can be written as

$$\left(\tilde{\varphi}_p\big|_{r=\mathcal{R}}, \tilde{\varphi}_q\big|_{r=\mathcal{R}}\right) = \oint_l \left[\tilde{\varphi}_p\big|_{r=\mathcal{R}}(z)\right]\left[\tilde{\varphi}_q^*\big|_{r=\mathcal{R}}(-z)\right]dl$$
$$= \oint_l \left[\tilde{\varphi}_p\big|_{r=\mathcal{R}}(z)\right]\left[\mathcal{PT}\,\tilde{\varphi}_q\big|_{r=\mathcal{R}}(z)\right]dl, \qquad (60)$$

where $l = 2\pi\mathcal{R}$ is a contour surrounding a cylindrical ferrite core. This relation may presume the absence of complex eigenvalues for $L$ modes. The phase acquired by the MDM fields orbitally rotating in a ferrite disk is $\Phi = k\pi$, where $k$ is an integer [86]. The relation (60) has different meanings for even and odd quantities $k$. For even quantities $k$, the edge waves show reciprocal phase behavior for propagation in both azimuthal directions. Contrarily, for odd quantities $k$, the edge waves propagate only in one direction of the azimuth coordinate. In a general form, the inner product (60) can be written as $\left(\tilde{\varphi}_p\big|_{r=\mathcal{R}}, \tilde{\varphi}_q\big|_{r=\mathcal{R}}\right) = (-1)^k \delta_{p,q}$, where $\delta_{p,q}$ is the Kronecker delta.

While a $PT$-symmetric Hamiltonian is, in general, not Hermitian, in the problem under consideration we can introduce a certain operator $\hat{\mathcal{C}}$, that the action of $\hat{\mathcal{C}}$ together with the $PT$ transformation will give the hermiticity condition and real-quantity energy eigenstates [90]. The operator $\hat{\mathcal{C}}$ is found from an assumption that it acts only on the boundary conditions of the $L$-mode spectral problem. This special differential operator has a form $\hat{\mathcal{C}} = i\frac{\mu_a}{\mathcal{R}}\left(\vec{\nabla}_\theta\right)_{r=\mathcal{R}}$, where $\left(\vec{\nabla}_\theta\right)_{r=\mathcal{R}}$ is the spinning-coordinate gradient. It means that, for a given direction of a bias field, operator $\hat{\mathcal{C}}$ acts only for a one-directional azimuth variation. The eigenfunctions of operator $\hat{\mathcal{C}}$ are double-valued border functions [34]. This operator allows performing the transformation from the natural boundary conditions of the $L$ modes, expressed by Eq. (7), to the essential boundary conditions of $G$ modes, which take the form of Eq. (58) [34, 92, 93]. Using operator $\hat{\mathcal{C}}$, we construct a new inner product structure for the boundary functions,

$$\left(\tilde{\varphi}_m\big|_{r=\mathcal{R}}, \tilde{\varphi}_n\big|_{r=\mathcal{R}}\right) = \oint_l \left[\tilde{\varphi}_m\big|_{r=\mathcal{R}}(z)\right]\left[\hat{\mathcal{C}}\mathcal{PT}\,\tilde{\varphi}_n\big|_{r=\mathcal{R}}(z)\right]dl, \qquad (61)$$

As a result, one has the energy eigenstate spectrum of MDM oscillations with topological phases accumulated by the double-valued border functions [34, 86]. The topological effects become apparent through the integral fluxes of the pseudoelectric fields. There are positive and negative fluxes corresponding to the clockwise and counterclockwise edge-function chiral rotations. For an observer in a laboratory frame, we have two oppositely directed anapole moments $\vec{a}^e$. This anapole moment is determined by the term $i\frac{\mu_a}{\mathcal{R}}\left(\frac{\partial \tilde{\varphi}}{\partial \theta}\right)_{r=\mathcal{R}^-}$ in Eq. (7). For a given direction of a bias magnetic field, we have two cases: $\vec{a}^e \cdot \vec{H}_0 > 0$ and $\vec{a}^e \cdot \vec{H}_0 > 0$.

In addition to the above consideration, other evident aspects prove the $PT$–symmetry in a geometrically symmetric structure with a MDM ferrite disk. It follows from the $PT$–symmetry of a scalar parameter characterizing properties of a ME field. At the MDM resonance, energy density of the ME-field structure is defined by the helicity factor $F^{(E)}$ [94]. There is the energy density of a quantized state of the near field originated from magnetization dynamics in a

quasi-2D ferrite disk. In a geometrically symmetric structure, the distribution of a helicity factor $F^{(E)}$, expressed by Eq. (22), is symmetric with respect to in-plane *x*, *y* coordinates and antisymmetric with respect to *z* coordinate [35, 43]. The helicity factor $F^{(E)}$ is a pseudoscalar. It changes a sign under inversions (also known as parity transformations). On the other hand, the factor $F^{(E)}$ changes its sign when one changes the direction of a bias magnetic field, that is, the direction of time (see Fig. 20). In a view of *PT*-symmetry of MDM oscillations (which is proven by an analyses of the MS-wave operator [86] and the time-space symmetry properties of the helicity factor $F^{(E)}$), the presence of an axial-vector orbital angular-momentum of a ferrite disk $\vec{\mathcal{L}}$ presumes also existence a certain polar vector directed parallel/antiparallel to the vector $\vec{\mathcal{L}}$. Such a polar vector exists. This is an anapole moment $\vec{a}^e$ originated from the $\pi$- flux resonances due to surface magnetic currents on a lateral surface of a ferrite disk [34, 95].

There can be different types of the microwave structures where *PT*-symmetry of MDMs in a ferrite disk is broken. The effect of the *PT*-symmetry breaking can be observed when, for example, a ferrite disk is placed in a rectangular waveguide non-symmetrically with respect to its walls. In such a case, the distribution of a helicity factor becomes assymetric with respect to *z* coordinate [35]. This is due to the role of boundary conditions on a metal surface. Inside a metal a helicity factor is zero. Fig. 21 shows schematically four cases of the helicity-factor distributions in a waveguide structure when two external parameters – a disk position on *z* axis and a direction of a bias magnetic field – change. At such a nonsymmetric disk position, the *PT* symmetry of MDM oscillations is broken. Other cases are related to microwave waveguiding structures with geometric nonsymmetry or the structures with inserted special chiral objects. As an important criterion for *PT*-symmetry breaking, there is the symmetry breaking of the helicity-factor distribution. Nonreciprocity effects, observed in the scattering matrix parameters of a microwave structure at MDM resonances, are also due to *PT*-symmetry breaking. Numerous studies of these properties for MDM structures, both numerical and experimental, are shown in Refs. [35, 65, 96, 97].

To a certain extent, the above questions on the *PT* symmetry breaking are akin to recent research in optics. For example, in publications [98 – 100], related to nonreciprocity in optics, it was shown that this effect takes place due to the *PT* symmetry breaking.

## VI. PREDICTION OF EXCEPTIONAL POINTS FOR A STRUCTURE OF MDMs IN THE CONTINUUM

MDM oscillations exhibit eigenstate chirality which is related to the Berry phase. The phase of the MDM wave function is not restored after one encircling around a lateral surface of the disk, but is changed by $\pi$. Only the second encircling restores the MS wave function including its phase [34]. There is the degeneracy of Hermitian operators (*G* modes), at which the eigenvalues coalesce while the eigenvectors remain different. The *L*-mode analysis [34, 89] shows that the eigenvector basis is skewed and the eigenvalues themselves are complex. Depending on a direction of a bias magnetic field, only one MDM state (clockwise or counterclockwise) dominates. In an analysis of MDM resonant structures, there are not only parameters controlling internal properties of the separated *L*-mode ferrite disk. There are also parameters by means of which the coupling strength between the MDM disk and the environment can be controlled. When we consider our microwave structures with an embedded MDM ferrite disk and inserted chiral objects, we also should consider the chirality of the eigenstates and the external-parameter chirality. In Ref. [96], the observations support the model which assigns a special role of the MDM chiral edge states in the unidirectional ME field multi-resonant tunneling. In Ref. [97], we have experimental observation of microwave chirality discrimination in liquid samples due to MDM eigenstate chirality. There is an evident correspondence between topological edge modes of MDM oscillations and "external" chirality of inserted objects.

For open resonant structures, an important aspect concerns a behavior of non-Hermitian systems at the spontaneous breaking *PT*–symmetry, marked by the exceptional point (EP). EPs are exhibited as a distinct class of spectral degeneracies, which are branch points in a 2D parameter space. The EP is the degeneracy intrinsic to non-Hermitian Hamiltonians at which two eigenvalues and corresponding eigenvectors coalesce. The wave function at an EP is a specific superposition of two configurations. The phase relation between the configurations is equivalent to a chirality [101 – 103]. The *PT*-symmetry breaking is related to chirality at the external-parameter loop. The sign of the chirality is defined via the direction of time. In the experiment in Ref. [102, 103], the positive direction of time is given by the decay of the eigenstates.

In our case, the time direction is given by the direction of a bias field. This is an external parameter of a type that can be controlled. In the prediction of exceptional points for a structure of MDMs in the continuum, we will use two tunable external parameters: (*i*) a bias magnetic field $H_0$ and (*ii*) the disk shift $d$ in a waveguide along $z$ axis, when the disk plane is parallel to waveguide walls. Evidently, if a MDM ferrite disk is placed in a waveguide symmetrically to its wall (z = 0), the direction of $z$ axis is arbitrary and either (positive or negative) time directions are the same. When the disk is shifted along $z$ axis, we can distinguish a definite direction of time. We chose the positive direction of time by an orientation of a bias magnetic field along $+z$ axis. Now, let a bias magnetic field is varied in a range $H_0^{(1)} < H_0 < H_0^{(2)}$. A MDM ferrite disk be shifted along $+z$ axis and its positions are defined as $d_1 \leq d \leq d_2$. The two external parameters compose a certain interaction parameter $\lambda(H_0, d)$.

We suppose that, similar to the structure shown in Fig. 8, the fields of a microwave continuum are the fields of a microwave cavity. Also, similar to the resonances shown in Figs. 15 and 16, we have two modes. Following the picture in Fig. 16, the modes are related to (*a*) the single-rotating-magnetic-dipole (SRMD) resonance; (*b*) double-rotating-magnetic-dipole (DRMD) resonance. By variation of the quantity of a bias magnetic field, we have an adjustable frequency difference $f_1 - f_2$ for these two modes. When a ferrite disk is shifted, with variation of quantity $d$, one observes variation of the helicity distribution. For the above resonances, we have variation of the field amplitudes and variation of the phases between the electric and magnetic fields [35, 43]. In the case of $f_1 = f_2$, there should be the possibility to adjust the phase difference between the modes (by variation of the disk shift along $z$ axis). One of the main evidence for existence of the EP is exhibition of chirality when one encircles this point in the space of $\lambda$ [101, 102]. When the EP is encircled, the eigenvectors pick up geometric phases. One of the evidence for this is the presence of the $\pi$ - flux resonances. There should exist a critical value $\lambda^{(EP)}\left(H_0^{(EP)}, d^{(EP)}\right)$, where the two modes coalesce.

The EP can be found by looking at the behavior of the real and imaginary parts of the eigenvalues. Such a basic analysis is beyond the frames of the present paper. It is a purpose of our future research. However, a qualitative analysis allows to predict encircling exceptional points along two paths on a parametric space created by a bias magnetic field and a disk shift. We predict four types of contours on such a 2D parametric space (see Fig. 22). When a ferrite disk is shifted along $+z$ axis at the positions defined as $d_1 < d < d_2$, two types of contours exist. For the range of positive magnetic field $H_0^{(1)} < H_0 < H_0^{(2)}$, we have an exceptional point **EP$_1$**$^{CW}$ inside the clockwise contour: A ➔ A′ ➔ B′ ➔ B ➔ A . For the range of negative magnetic field $-H_0^{(1)} > -H_0 > -H_0^{(2)}$, an exceptional point **EP$_2$**$^{CCW}$ is inside the counterclockwise contour: $C \to C' \to D' \to D \to C$. From Fig. 21, one can see that a simultaneous change of a sign of the bias magnetic field $H_0$ and a sign of the disk shift $d$ in a waveguide gives symmetrically the same pictures of the helicity distribution. It means that there exist two other contours on our 2D parametric space. When a ferrite disk is shifted along $-z$ axis at the positions $-d_1 > -d > -d_2$ and the range of positive magnetic field is $H_0^{(1)} < H_0 < H_0^{(2)}$, we have an exceptional point $\mathbf{EP_3^{CCW}}$ inside the counterclockwise contour: $E \to E' \to F' \to F \to E$. At the positions $-d_1 > -d > -d_2$ and the range of negative magnetic field $-H_0^{(1)} > -H_0 > -H_0^{(2)}$, an exceptional point $\mathbf{EP_4^{CW}}$ is inside the counterclockwise contour: $G \to G' \to H' \to H \to G$. It is evident that $\mathbf{EP_1^{CW}}$ is the same as $\mathbf{EP_4^{CW}}$ and $\mathbf{EP_2^{CCW}}$ is the same as $\mathbf{EP_3^{CCW}}$. We can see that in a case of a waveguide with a non-symmetrically embedded ferrite disk, the ME near-field distribution becomes not *PT* symmetrical. However, for the entire microwave structure we have *PT* symmetry together with chiral symmetry of contours on a 2D parametric space.

To verify the above prediction of the EPs in a microwave waveguide with a non-symmetrically embedded ferrite disk, we will do the following qualitative analysis of the problem. Energy density of the ME near-field structure is defined by the helicity-factor density $F^{(E)}$ [94]. We have the regions with the positive and negative ME-field densities, which are characterized by the positive and negative helicity-factor densities, $F^{(E)(+)}$ and $F^{(E)(-)}$, respectively. We define the positive helicity (the positive ME energy) as an integral of the ME-field density over the entire near-field vacuum region $\mathbb{V}^{(+)}$ with the helicity-factor density $F^{(E)(+)}$:

$$\mathbb{H}^{(+)} = \int_{\mathbb{V}^{(+)}} F^{(E)(+)} d\mathbb{V}. \tag{62}$$

Similarly, the negative helicity (the negative ME energy) is defined as an integral of the ME-field density over the entire near-field vacuum region $\mathbb{V}^{(-)}$ with the helicity-factor density $F^{(E)(-)}$:

$$\mathbb{H}^{(-)} = \int_{\mathbb{V}^{(-)}} F^{(E)(-)} d\mathbb{V}. \tag{63}$$

In a case of *PT* symmetrical ME near-field distribution, $\mathbb{H}^{(+)} = -\mathbb{H}^{(-)}$. So, the total helicity (the total ME energy) of the entire near-field vacuum region surrounding a ferrite disk $\mathbb{V} = \mathbb{V}^{(+)} + \mathbb{V}^{(-)}$ is equal to zero [94]:

$$\mathbb{H} = \mathbb{H}^{(+)} + \mathbb{H}^{(-)} = \int_{\mathbb{V}} \left( F^{(E)(+)} + F^{(E)(-)} \right) d\mathbb{V} = 0. \tag{64}$$

If, however, the *PT* symmetry is broken (in a case, for example, when a ferrite disk is shifted in a waveguide along $z$ axis), we have $\left|\mathbb{H}^{(+)}\right| \neq \left|\mathbb{H}^{(-)}\right|$. It means that we may have predominance of the positive or negative ME energy. In particular, in Fig. 21 (*a*) we have predominance of the negative ME energy, while in Fig. 21 (*b*), there is predominance of the positive ME energy. On a metal wall, the helicity-factor density (and so, the ME-energy density) is zero. Where does the excess of ME energy go?

The ME fields are quantized state of the near fields originated from the magnetization dynamics in a quasi-2D ferrite disk. There are pseudoscalars, which appear as the fields of axion electrodynamics. The coupling between an axion field and the electromagnetic field results in modification of the electric charge and current densities in electrodynamics equations. Such modified sources of electromagnetic fields were observed on metal walls of a microwave waveguide with an embedded MDM ferrite disk [36, 82, 86]. It is worth noting here that our ME-field structures are fundamentally different from the ME fields considered in Ref. [104]. In Ref. [104] it has been proposed, hypothetically, that local ME fields can be realized due to the interference of several plane waves. In such a case, however, the question, how one can create a pseudoscalar field from interaction of regular electromagnetic waves, remains open.

Absorption of ME energy takes place only at the *PT*-symmetry breaking. When, in particular, the *PT*-symmetry-breaking-behavior is realized by shifting a ferrite disk in a waveguide along $z$ axis, the excess of ME energy (positive or negative) leads to appearance of topological charges and currents induced on the upper and lower walls of a waveguide. In Refs. [36, 82], it was shown numerically that in a structure with an embedded ferrite disk, surface electric currents on a waveguide wall are the right-handed and left-handed flat spirals. For the *PT*-symmetric structure, we have chiral symmetry of such currents on the upper and lower waveguide walls. In a case of the *PT*-symmetry breaking, spontaneous chiral symmetry breaking occurs and the net chirality of the surface currents can fall into either the left-handed or the right-handed regime. This results in absorption of ME energy for MDMs.

Let in the MDM oscillation spectra, there be two Hermitian-Hamiltonian $G$ modes with energies $E_1$ and $E_2$, which are sufficiently close one to another. We assume that coupling between these modes is reciprocal and is characterized by a real coefficient $\kappa$. For an open structure with *PT*–symmetry breaking, the a non-Hermitian Hamiltonian of the two-state system is expressed as

$$\mathcal{H} = \begin{pmatrix} E_1 \pm i\gamma_1 & \kappa \\ \kappa & E_2 \pm i\gamma_2 \end{pmatrix}. \quad (65)$$

Real quantities $\gamma_1$ and $\gamma_2$ characterize the absorption rates. For the two modes, the absorption is due to the radiation inside a waveguide continuum caused be the currents induced on the waveguide walls of the *PT*-symmetry-breaking structure. The $\pm$ signs in Eq. (65) characterize different kinds of the ME-energy excess (positive or negative). When the signs of imaginary components in Eq. (65) are the same, an analytical treatment of encircling of EPs can be similar to that used in Ref. [103].

As we discussed above, due to the topological action of the azimuthally unidirectional transport of energy in a MDM-resonance ferrite sample there should exist the opposite topological reaction (opposite azimuthally unidirectional transport of energy) on a metal screen placed near this sample. Conservation of the orbital angular momentum, related to the power-flow circulation, has been proven in a *PT*-symmetrical microwave structure with an embedded ferrite disk [36]. It was also discussed that a vacuum-induced Casimir torque allows for torque transmission between the ferrite disk and metal wall avoiding any direct contact between them. For the *PT*-symmetrical microwave structure, the Casimir energy is the same above and below a ferrite disk. When the *PT* is broken, the Casimir energy above and below a ferrite disk is not the same. The orbital angular momentum is not conserved now. All this means that together with chirality of the MDM ferrite disk we should observe chirality of the microwave continuum. At the MDM resonances, the eigenvector basis of continuum becomes skewed and the eigenvalues of continuum become complex. This appears clearer from the fact that in microwave structures with MDM ferrite particles we can observe path-dependent interference [82]. In general, we have a case of non-integrable systems. Following Ref. [105], there is the possibility to introduce two topological numbers: (*i*) related to the Berry-curvature chirality of the eigenstates and (*ii*) the exceptional-point chirality in the external-parameter space.

It is worth noting that when a MDM ferrite disk is placed in a rectangular waveguide symmetrically with respect to its walls, *PT* symmetry is valid when one can neglect asymmetry in the ME-field structure due to the wave propagation direction in a waveguide. This is possible in a case of a quasi-2D microwave structure with a "thin" waveguide (see Fig. 4), analyzed in Ref. [36]. In a microwave structure with a "thick" waveguide, there is an evident asymmetry in the distribution of a helicity factor with respect to *x*, *y* coordinates [35, 82, 83]. Such asymmetry due to the wave propagation direction in a waveguide can lead to unique topological effects. In Ref. [82], it was shown that in a microwave structure of two ferrite disks placed on a waveguide axis symmetrically to the waveveguide walls, the coupling of two identical MDM resonances is nonreciprocal.

## VII. DISCUSSION: ON THE COMPLEX-WAVE INTERACTION BETWEEN EM AND ME PHOTONS

In Ref. [106], an optical system that combines a transverse spin and an evanescent electromagnetic wave in semi-space geometry, has been proposed. In a theoretical model, it is assumed that in such surface modes with strong spin-momentum locking the real and imaginary EM wave vectors are mutually perpendicular. Nevertheless, it becomes clear that a slight deviation of the spin vector from the normal to the surface-wave propagation direction, will lead to appearance of a leaky wave, which is treated, mathematically, as a guided complex wave.

The question of a complex-wave behavior is very important in our case of a microwave structure with an embedded ferrite disk. In the near-field region, ME photons, originated from a quasi-2D ferrite disk, are characterized by the fields orbitally rotating in a plane parallel to the disk plane and decaying along the disk axis. Such a field structure can be described by the MS-wave function with mutually perpendicular a real wave vector (in a plane parallel to the disk plane) and an imaginary wave vector (directed along the disk axis). When analyzing in Ref. [36] the electric current induced on a metal wall in a waveguide with an embedded MDM ferrite disk, we found the vortex structure of this current, we can also assume that the field near to this wall has mutually perpendicular real (in a plane parallel to the metal wall) and imaginary (along a normal to the wall) wave vectors. However, when we take into account the waveguide thickness, that is when we consider a 3D microwave structure, the real and imaginary wave vectors of the ME-field structure will have co-parallel components. It means that in the near-field region, the ME field is viewed as a complex-wave field. This complex-wave behavior is observed together with the spin-orbit coupling in the ME-field structure. It becomes clear that together with localized and quantized ME-field complex waves, we have to presume the presence of the complex-wave continuum in a microwave waveguide.

In the above discussions, regarding experimental results obtained in Ref. [32] (and shown in the present paper in Fig. 8), we stated that for the microwave photon propagating in a waveguide at a constant frequency, the energy extracted from the magnetic energy of a ferrite disk at the MDM resonances cannot be a quantized quantity. So, the question of which microwave continuum we have when observing discrete states of MDM resonance in a microwave cavity with a constant frequency, remains open. To answer this question, we should consider the spectrum of so-called complex modes in lossless microwave waveguides. In a closed lossless waveguide, there can be a finite number of real positive eigenvalues and an infinite number of complex conjugate pairs of eigenvalues [107, 108]. Following the parameter circulation in Fig. 12, we can see that for a given frequency $\omega$, the input impedance of the cavity passes through complex quantities for every separate MDM. Assuming the presence of complex waves in the waveguide spectrum we should conclude that in Fig. 13 not only impedance $Z_L$ is a complex quantity but also a characteristic impedance of a lossless waveguide $Z_0$ is a complex quantity. Generally, we have to take into account the possibility of interaction between discrete-state complex-wave MDMs and the microwave complex-wave continuum.

Let us consider general relations for mode orthogonality in a lossless regular waveguide assuming the presence of complex waves in the waveguide spectrum. We write homogeneous Maxwell equations in an operator form:

$$M\vec{U} = 0, \tag{66}$$

where $M$ is the Maxwell operator:

$$M = \begin{pmatrix} i\omega\varepsilon_0 & -\vec{\nabla}\times \\ \vec{\nabla}\times & i\omega\mu_0 \end{pmatrix} \tag{67}$$

and $\vec{U}$ is a vector function of the fields:

$$\vec{U} = \begin{pmatrix} \vec{E} \\ \vec{H} \end{pmatrix}. \tag{68}$$

In a regular waveguide, the solution of Eq. (66) for electromagnetic wave propagating along $z$ axis is presented based on the mode expansion:

$$\vec{U} = \sum_m A_m \hat{\vec{U}}_m e^{-\gamma_m z}, \tag{69}$$

where $\gamma_m = \alpha_m + i\beta_m$ is a propagation constant of a normal mode $m$ and $\hat{\vec{U}}_m$ is a membrane functions on the waveguide cross section:

$$\hat{\vec{U}}_m(x, y) = \begin{pmatrix} \hat{\vec{E}}_m(x, y) \\ \hat{\vec{H}}_m(x, y) \end{pmatrix}. \tag{70}$$

For homogeneous Maxwell equations, we can write:

$$M_\perp \hat{\vec{U}}_m = \gamma_m Q \hat{\vec{U}}_m, \tag{71}$$

where $M_\perp$ is differential-matrix operator similar to the operator $M$ but operating over cross section coordinate of a waveguide and

$$Q = \begin{pmatrix} 0 & -\vec{e}_z \times \\ \vec{e}_z \times & 0 \end{pmatrix}. \tag{72}$$

Here $\vec{e}_z$ is the unit vector directed along $z$ axis.

We juxtapose now the differential problem described by Eq. (71) with the conjugated problem. For the conjugated problem, we have:

$$\tilde{M}_\perp \hat{\vec{V}}_{\tilde{m}} = \tilde{\gamma}_{\tilde{m}} Q \hat{\vec{V}}_{\tilde{m}}, \tag{73}$$

where $\gamma_{\tilde{m}} = \alpha_{\tilde{m}} + i\beta_{\tilde{m}}$ is a propagation constant of a normal mode $\tilde{m}$ and $\hat{\vec{V}}_{\tilde{m}}$ is a vector function of the membrane-function fields of the conjugated problem:

$$\hat{\vec{V}}_{\tilde{m}}(x,y) = \begin{pmatrix} \hat{\vec{E}}_{\tilde{m}}(x,y) \\ \hat{\vec{H}}_{\tilde{m}}(x,y) \end{pmatrix}. \tag{74}$$

A form of the conjugate operator $\tilde{M}_\perp$ is defined from integration by parts [92]:

$$\int_S \left( M_\perp \hat{\vec{U}}_m \right) \hat{\vec{V}}_{\tilde{m}}^* dS = \int_S \hat{\vec{U}}_m \left( \tilde{M}_\perp \hat{\vec{V}}_{\tilde{m}} \right)^* dS + \oint_\mathcal{L} \hat{\vec{U}}_m \left( R \hat{\vec{V}}_{\tilde{m}} \right)^* d\mathcal{L}, \tag{75}$$

where $S$ is the waveguide cross-section area, $\mathcal{L}$ is the contour surrounding $S$. The operator $R$ has a form

$$R = \begin{pmatrix} 0 & \vec{n} \times \\ -\vec{n} \times & 0 \end{pmatrix}, \tag{76}$$

where $\vec{n}$ is the unit vector directed along the external normal to contour $\mathcal{L}$. For a lossless waveguide and homogeneous boundary conditions, operator $\tilde{M}_\perp$ is the same as operator $M_\perp$. As a result, one has the orthonormality condition

$$\left( \gamma_m + \gamma_{\tilde{m}}^* \right) \int_S \left( Q \hat{\vec{U}}_m \right) \hat{\vec{V}}_{\tilde{m}}^* dS = 0. \tag{77}$$

Eq. (68) is correct for a general case of complex waves in a lossless waveguide. In this case, we have biothogonality for conjugate modes. The conjugate modes are $m$ and $\tilde{m}$ are orthogonal when $\gamma_m + \gamma_{\tilde{m}}^* \neq 0$. Otherwise, these modes constitute a norm:

$$N_{m\tilde{m}} = \int_S \left( Q \hat{\vec{U}}_m \right) \hat{\vec{V}}_{\tilde{m}}^* dS = \int_S \left( \hat{\vec{E}}_m \times \hat{\vec{H}}_{\tilde{m}}^* + \hat{\vec{E}}_{\tilde{m}}^* \times \hat{\vec{H}}_m \right) \vec{e}_z dS. \tag{78}$$

In the spectrum of a lossless waveguide, there are four complex modes: $\gamma_m = \alpha_m + i\beta_m$, $\gamma_{\tilde{m}} = -\alpha_m + i\beta_m$, $\gamma_{-m} = -\alpha_m - i\beta_m$, and $\gamma_{-\tilde{m}} = \alpha_m - i\beta_m$, which exist in fours: $\gamma_m = -\gamma_{\tilde{m}}^*$, $\gamma_m = \gamma_{-\tilde{m}}^*$, $\gamma_{-m} = -\gamma_{-\tilde{m}}^*$, $\gamma_{-m} = \gamma_{\tilde{m}}^*$. Positions of these complex wave numbers are shown in Fig. 23. An interaction of modes with $\gamma_m$ and $\gamma_{\tilde{m}}$ or $\gamma_{-m}$ and $\gamma_{-\tilde{m}}$ cause appearance of active power flows; meanwhile an interaction of modes with $\gamma_m$ and $\gamma_{-\tilde{m}}$ or $\gamma_{-m}$ and $\gamma_{\tilde{m}}$ lead to appearance of reactive power flows. The pairs of modes which realize the carrying over of energy are characterised by the same direction of phase velocity and different sign of amplitude changing [107, 108].

In a particular case of a propagating-wave behavior, we have $\gamma_m = \gamma_{\tilde{m}} = i\beta_m$. For propagating modes $\hat{\vec{E}}_m = \hat{\vec{E}}_{\tilde{m}}$, $\hat{\vec{H}}_m = \hat{\vec{H}}_{\tilde{m}}$, the electric and magnetic fields oscillate in phase. So, the norm (78) is a real quantity. On the other hand, in a particular case of an evanescent wave, the electric-field oscillations are $90°$ phase shifted in time with respect to the magnetic-field oscillations. So, the norm defined by Eq. (78) is an imaginary quantity. At such an evanescent-wave behavior no transmission of energy occurs. However, one can have power transmission (tunneling) at certain boundary conditions of a section of a below-cutoff waveguide. Evanescent waves are non-local waves. So, normalization is non-local as well. To get transmission in a below cut-off section we have to take into account the load conditions. Let the incident field be the positive decaying mode and the reflected field be the negative decaying mode (see Fig. 24). Such reflected and incident fields can be related through a certain load-dependent reflection coefficient [109]. Fig. 24 shows the complex-plane combination of incident and reflected fields that will exist at the end of a below-cutoff waveguide section. The port 1 is considered as a source. The section is terminated in a certain load at port 2. The above expressions involve both normal modes, with fields decaying in positive and negative, respectively. We consider $\hat{\vec{E}}_m \equiv \vec{E}_i$, $\hat{\vec{E}}_{\tilde{m}} \equiv \vec{E}_r$ and $\hat{\vec{H}}_m \equiv \vec{H}_i$, $\hat{\vec{H}}_{\tilde{m}} \equiv \vec{H}_r$, where $\vec{E}_i$, $\vec{E}_r$ $\vec{H}_i$, $\vec{H}_r$ are the incident and reflected electric and magnetic fields, respectively. We also assume that both incident and reflected electric and magnetic fields are linearly polarized in space and that $\gamma_m = -\gamma_{\tilde{m}} = \alpha_m$. When, due to a load in port 2, the electric- and magnetic-field oscillations of modes $m$ and $\tilde{m}$ become $\pm 90°$ phase shifted in time, the total electric $\vec{E}_i + \vec{E}_r$ and total magnetic $\vec{H}_i + \vec{H}_r$ fields are in phase and the norm (78) is a real quantity. So, an interaction of modes with $\gamma_m$ and $\gamma_{\tilde{m}}$ causes appearance of active power flows through a below-cutoff waveguide section. This effect is similar to the tunneling effect in quantum mechanics structures. The two orthogonal basis states $|1\rangle \equiv \hat{\vec{U}}_m$ and $|2\rangle \equiv \hat{\vec{V}}_{\tilde{m}}$ oscillate in time if they are superimposed according to relation $|1\rangle \pm i|2\rangle \equiv \hat{\vec{U}}_m + i\hat{\vec{V}}_{\tilde{m}}$. This coalescent state is a chiral state. As an interaction parameter of a system – a below-cutoff waveguide section – there is a boundary condition at port 2. The pairwise coalescence occurs only at a certain boundary condition at port 2 that is at a certain value of an interaction parameter.

Now, we come back to the structure shown in Fig. 8. Following the above analysis, we suppose that the port 1 is the cavity input. The localized region in a waveguide where a MDM ferrite disk is situated, is the port 2 with a certain load. We also suppose that in a waveguide section between ports 1 and 2 the complex waves can be observed. Fig. 25 shows a possible region of existence of such complex waveguide modes on the $\omega - k$ diagram, where $k$ is a real wavenumber. The phases of the MDM bound states are strongly determined phases. We can assume that for every MDM there are certain quantities of a bias magnetic field that provide $\pm 90°$ phase shift in time for the waveguide fields at the region of a ferrite disk, that is the region of the port 2. This exactly corresponds to points 1, 2, 3, … in Figs. 8 and 12. Discretization of the microwave response is only due to this complex-wave behavior of waveguide modes. The conditions for normalization of these complex-wave modes is illustrated in Fig. 23.

How can we view a general picture of scattering of these complex waveguide modes by the resonant MDM particle? We suppose, initially, that a magnonic-resonance in a small ferrite particle is deprived of any orbital rotations (in other words, there are no vortices in the resonant modes). There is a resonant magnetic particle dual to plasmonic-resonance subwavelength particle. In this case, the scattering of propagating EM waves is well described by Mie theory. The particle near field is a decaying field described by Laplace equation. Now, we take into account that the quasistatic fields of this magnonic-resonance particle are orbitally rotating fields. Because of such a behavior, we have strongly determined localized phases. At the MDM resonances, this gives a "proper" scattering for evanescent EM waves. All this means that at MDM resonances, we observe scattering of waveguide complex-wave modes. In other words, we have an interaction of MDM discrete states with the complex-wave microwave continuum.

Till now, we do not have direct experimental and/or numerical proves of the proposed model of the complex-wave interaction between EM and ME photons. Nevertheless, some unique topological-phase properties shown in Ref. [82] should deserve our special attention regarding this aspect. It was shown that in a microwave structure of two small ferrite disks placed on a waveguide axis symmetrically to the waveguide walls [82], the frequency split of the interacting MDM resonances is extremely narrow and (what is especially unique) is quite independent on the distance between the disks. Since this distance varied from the near-field to far-field regions of the waveguide mode, the coupling between MDM disks is not due to the propagating EM wave, but because of the presence of the evanescent part of the complex EM wave. One of specific properties of evanescent and tunneling modes is that they are non-local. For EM waves, the MDM disks appear as localized-phase singularities and interaction between the disks is due to the microwave complex-wave continuum.

## VIII. CONCLUSION

In small ferromagnetic-resonance samples, macroscopic quantum coherence can be observed. Long range magnetic dipole-dipole correlation can be treated in terms of collective excitations of the system. In a case of a quasi-2D ferrite disk, the quantized forms of these collective matter oscillations – the MDM magnons – were found to be quasiparticles with both wave-like and particle-like behaviors, as expected for quantum excitations.

In an electromagnetically subwavelength ferrite sample one neglects a time variation of electric energy in comparison with a time variation of magnetic energy. In this case, the Faraday-law equation is incompatible with the spectral solutions for magnetic oscillations. It appears, however, that in a case of MDM oscillations in a quasi-2D ferrite disk, the Faraday equation plays an essential role. In a ferrite-disk sample, the magnetization has both the spin and orbital rotations. There is the spin-orbit interaction between these angular momenta. This results in unique properties when the lines of the electric field as well the lines of the polarization in a sample are "frozen" in the lines of magnetization. In such a case, the Faraday law is not in contradiction with the magnetostatic equations describing the MDM-oscillation spectra.

The MDMs in a ferrite disk are characterized by the pseudoscalar magnetization helicity parameter which gives evidence for the presence of two coupled and mutually parallel currents – the electric and magnetic ones – in a localized region of a microwave structure. The near fields of a MDM sample are so-called ME fields. The magnetization helicity parameter can be considered as a certain source which defines the helicity properties of ME fields. The ME fields, being originated from magnetization dynamics at MDM resonances, appear as the pseudoscalar axionlike fields. Whenever the pseudoscalar axionlike fields, is introduced in the electromagnetic theory, the dual symmetry is spontaneously and explicitly broken. This results in non-trivial coupling between pseudoscalar quasistatic ME fields and the EM fields in microwave structures with an embedded MDM ferrite disk.

Unique properties of interaction of MDMs with a metal screen become more evident when one analyzes the angular-momentum balance conditions for MDM oscillations in a ferrite disk in a view of the ME-EM field coupling in a microwave waveguide. MDMs in a quasi-2D ferrite disk are microwave energy-eigenstate oscillations with topologically distinct structures of rotating fields and unidirectional power-flow circulations. Due to the topological action of the azimuthally unidirectional transport of energy in a MDM-resonance ferrite sample there exists the opposite topological reaction (opposite azimuthally unidirectional transport of energy) on a metal screen placed near this sample. This effect is called topological Lenz's effect. In a microwave structure with an embedded ferrite disk, an orbital angular momentum, related to the power-flow circulation, must be conserved in the process. Thus, if power-flow circulation is pushed in one direction in a ferrite disk, then the power-flow circulation on metal walls to be pushed in the other direction by the same torque at the same time. A vacuum-induced Casimir torque allows for torque transmission between the ferrite disk and metal wall avoiding any direct contact between them.

The fact that magnetization dynamics in a quasi-2D ferrite disk have an impact on the energy quantization of the fields in a microwave cavity, was confirmed experimentally. Sharp multiresonance oscillations, observed experimentally in microwave structures with an embedded quasi-2D ferrite disk, are related to magnetization dynamics in the sample. This dynamics have an impact on the phenomena connected with the quantized energy fluctuation. Inside the YIG disk, we have the torque exerting on the electric polarization due to the magnetization dynamics. Because of the effective magnetic charges on a ferrite-disk planes, the demagnetizing magnetic field is reduced. It means that the DC magnetization of a ferrite disk is reduced as well. At the MDM resonances, we observe quantization of the DC magnetization of a ferrite disk. It is worth noting also that at the MDM resonance, a quasi-2D disk is manifested as a ME particle with two DC moments directed along the disk axis: (*i*) a DC magnetic moment (due to saturation magnetization) and (*ii*) a DC electric moment – the anapole moment.

In the paper, we proved *PT*-symmetry of MDMs in a ferrite disk and analyzed the conditions of the *PT*-symmetry breaking. When analyzing the scattering of EM waves by MDM disks in microwave waveguides and energy quantization of the field in a microwave cavity, we dwelled on some basic problems of magnon-photon interaction and bound states in the microwave continuum. For a *PT*-symmetrical structure, one of the features attributed to the BICs is a strong resonance field enhancement. In the case of MDM oscillations in a microwave-field continuum, such a strong field enhancement was confirmed in numerous numerical studies. In microwave-cavity structures with embedded small ferrite disks, the effect of Fano resonance collapse was shown at variation of an external parameter – a bias magnetic field. When a *PT* symmetry of a microwave structure with an embedded MDM ferrite disk is broken, one can predict existence of certain branch points – the exceptional points – in a 2D parameter space of the structure.

To answer the question of which microwave continuum we have when observing discrete states of MDM resonance in a microwave cavity with a constant frequency, we take into consideration the spectrum of so-called complex modes in lossless microwave waveguides. The near fields of a MDM-resonance particle are orbitally rotating fields. Because of such a behavior, we have strongly determined localized phases. For incident EM waves, this particle appears as a localized phase singularity. As an important problem, we discussed the possibility of interaction between discrete-state complex-wave MDMs and the microwave complex-wave continuum.

Finally, we have to note that interdisciplinary studies of EM-field chirality and magnetism is a topical subject in optics. The magnetic dipole precession has neither left-handed nor right-handed quality, that is to say, no chirality [20, 22]. The synthesis of chiral and magnetic properties in molecular structures [110] allow the observation of strong magneto–chiral dichroism, where unpolarized light is absorbed differently for parallel and antiparallel propagation with respect to an applied magnetic field. If the chiral medium itself is ferromagnetic, the large internal magnetic field should provide a significant boost to such phenomena compared with para- and diamagnetic media [111, 112]. While these magneto–chiral effects are related to specific local properties of materials, in our study, an interplay of the chiral and magnetic phenomenon is exhibited in a very different and unique manner. This phenomenon is due to topological properties of MDM oscillations in a quasi-2D ferrite disk together with special properties of the microwave continuum – the external microwave structure.

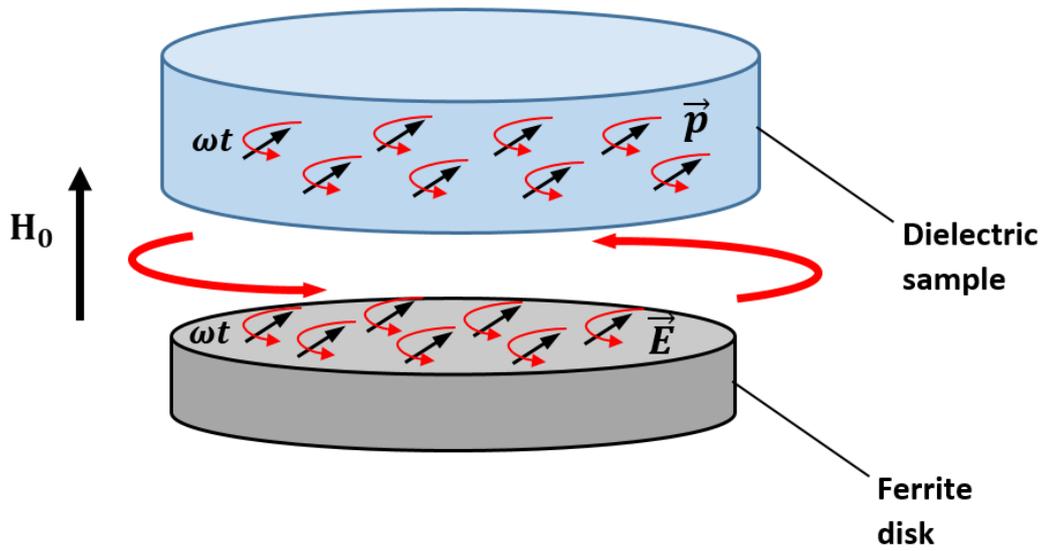

FIG. 1. The fields in a composed ferrite-dielectric structure at MDM resonances. Due to magnetization dynamics, an electric field inside a ferrite disk has both spin and orbital angular momentums. Electric dipoles induced in a dielectric sample precess and accomplish an orbital rotation. A bias magnetic field is directed normally to the disk plane. For opposite directions of a bias magnetic field, one has opposite rotations, both orbital and spin, of the fields inside a ferrite and a dielectric.

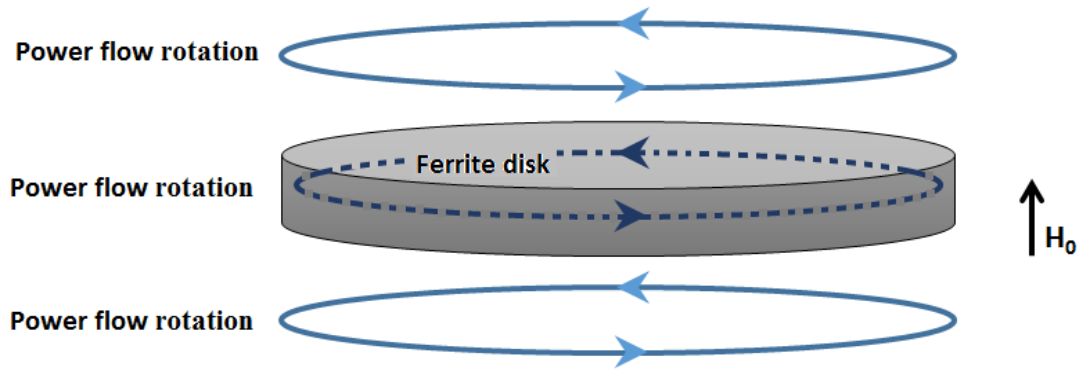

Fig. 2. An orbital angular momentum of a ferrite disk at a MDM resonance. For a given direction of a bias magnetic field, the power-flow circulations are the same inside a ferrite and in the vacuum near-field regions above and below the ferrite disk.

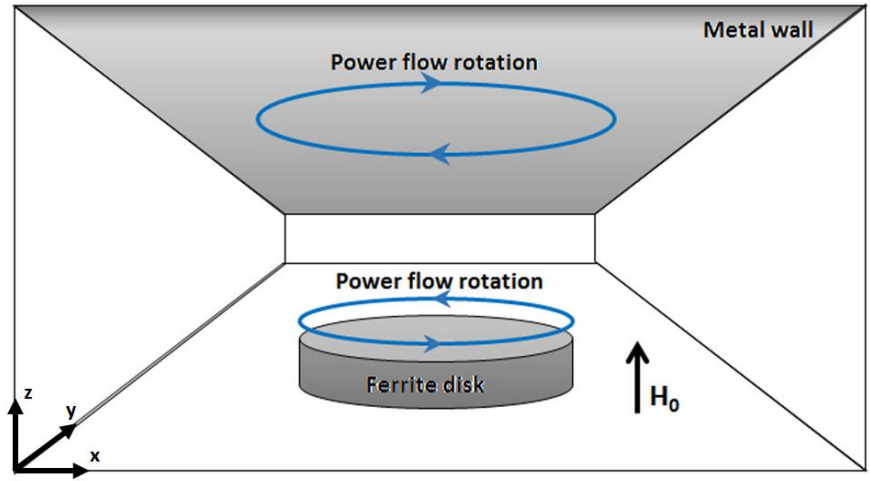

Fig. 3. The ME-EM field interaction in a microwave waveguide. Evidence for the angular momentum balance conditions at a given direction of a bias magnetic field. In a view along $z$ axis, there are opposite power flow circulations in a ferrite disk and on a metal surface.

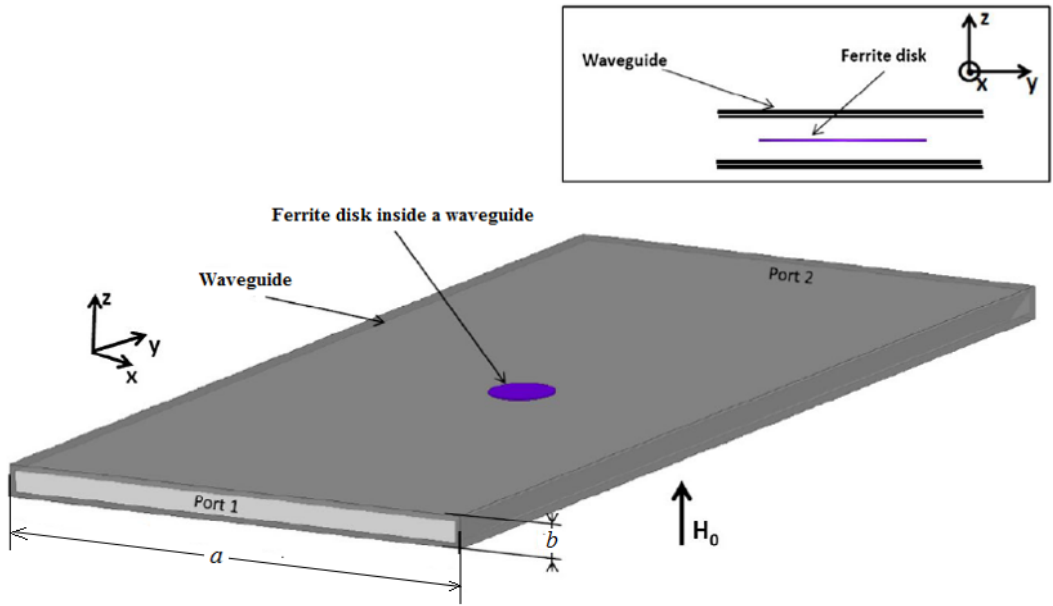

Fig. 4. A quasi-2D model of a thin rectangular waveguide with an embedded thin-film ferrite disk. An insert shows the disk position inside a waveguide.

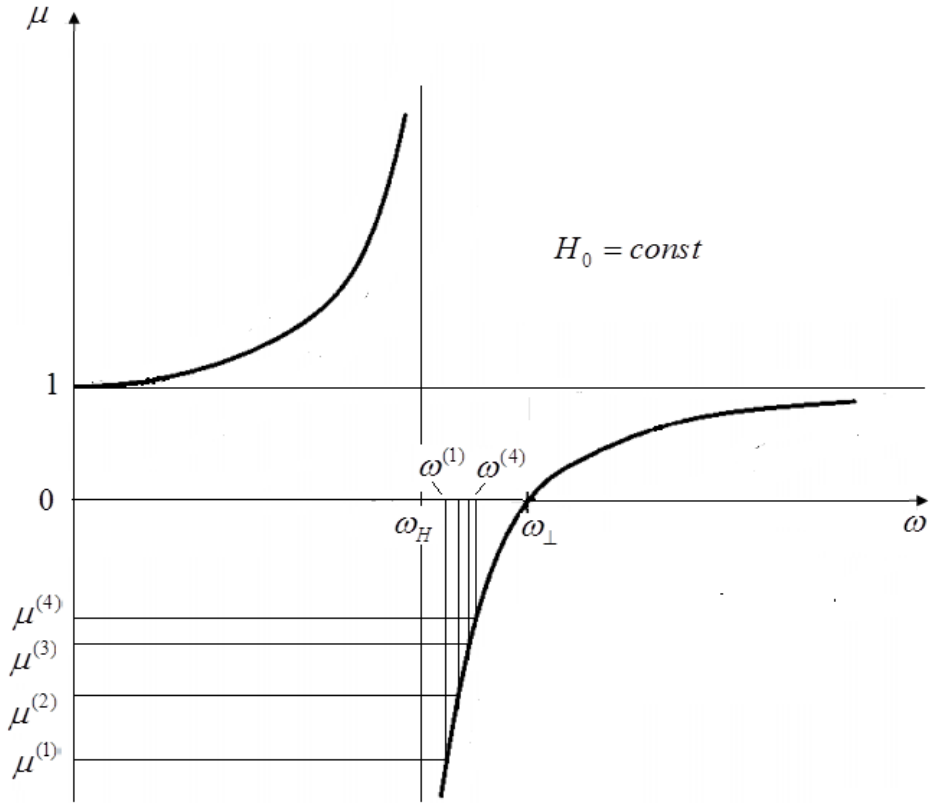

Fig. 5. FMR diagonal component of the permeability tensor $\mu$ versus frequency at a constant bias magnetic field. The discrete quantities, shown for the first four MDM resonances, are in the region where $\mu < 0$ $(\omega_H < \omega < \omega_\perp)$.

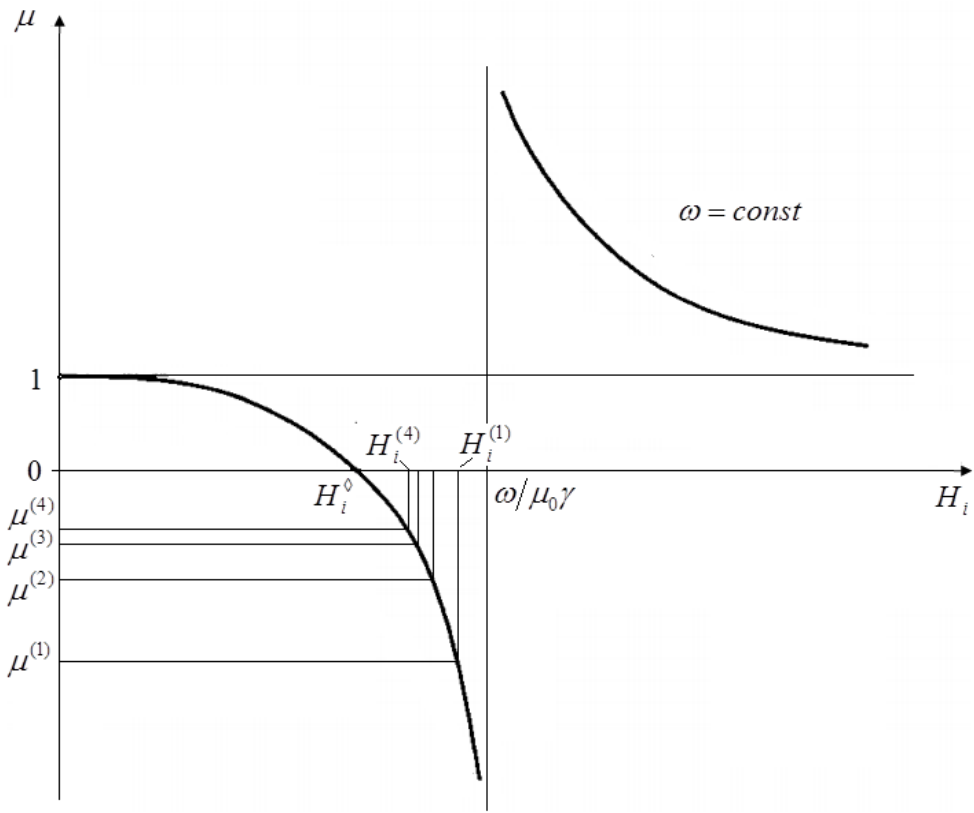

Fig. 6. FMR diagonal component of the permeability tensor versus a DC internal magnetic field at a constant frequency. The discrete quantities, shown for the first four MDM resonances, are in the region where $\mu < 0$ $\left( H_i^\diamond < H_i < \omega/\mu_0 \gamma \right)$.

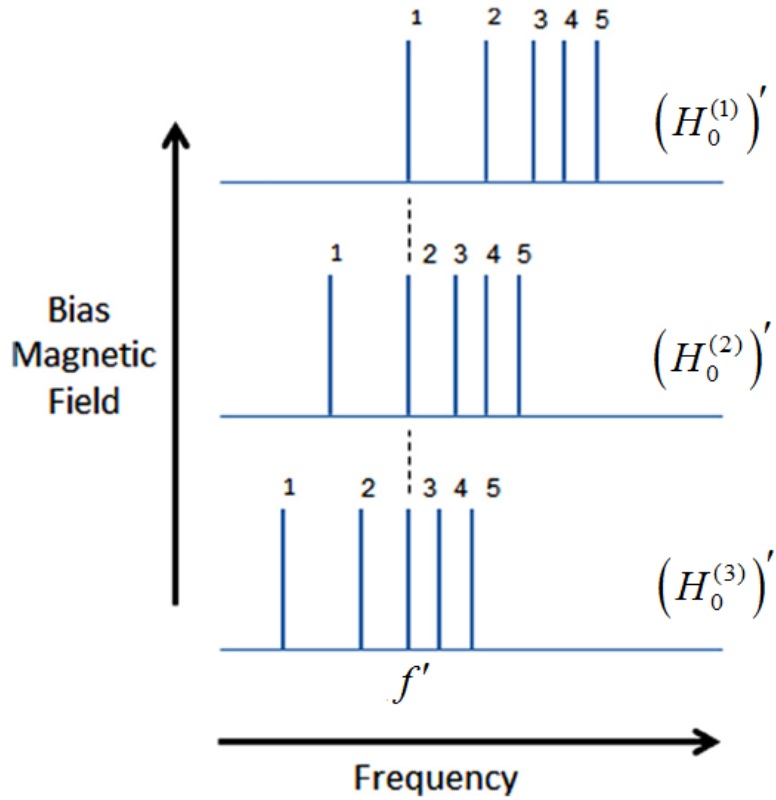

Fig. 7. Correlation between two mechanisms of the MDM energy quantization: Quantization by signal frequency and quantization by a bias magnetic field.

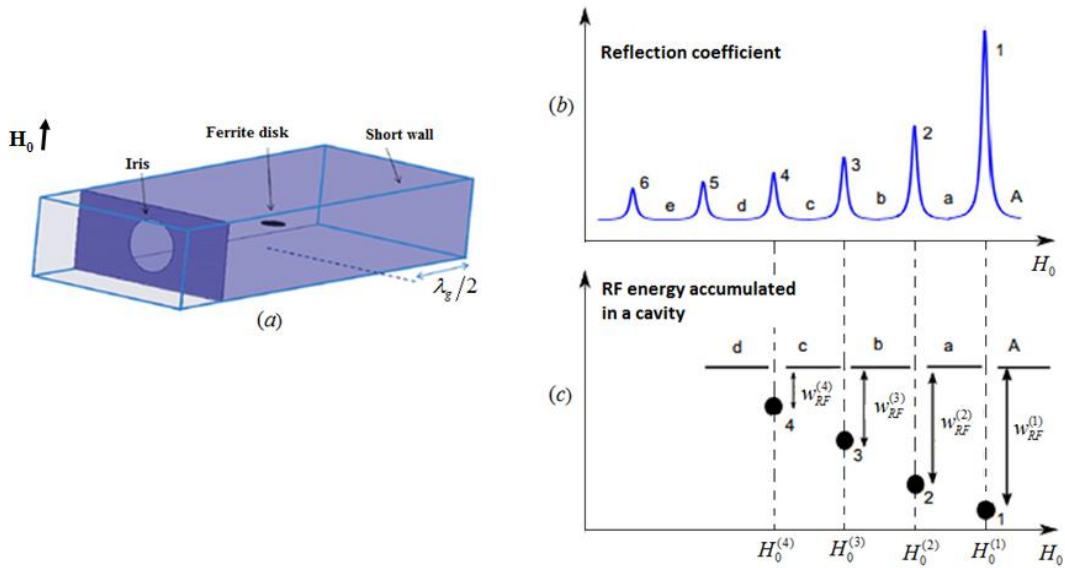

Fig. 8. Relationship between quantized states of microwave energy in a cavity and magnetic energy in a ferrite disk. (*a*) A structure of a rectangular waveguide cavity with a normally magnetized ferrite-disk sample. (*b*) A typical multiresonance spectrum of modulus of the reflection coefficient. (*c*) Microwave energy accumulated in a cavity; $w_{RF}^{(n)}$ are jumps of electromagnetic energy at MDM resonances.

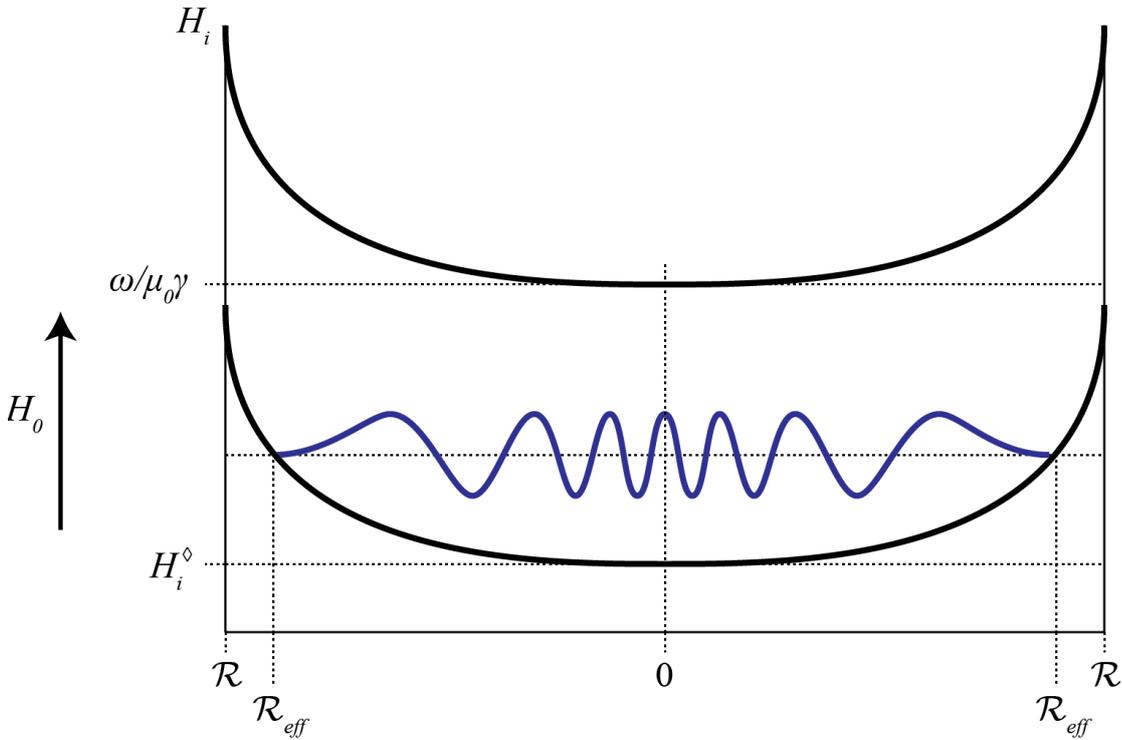

Fig. 9. A model showing a radial distribution of the MS-potential wave function based on the Bohr–Sommerfeld quantization rule. The region of an internal magnetic field where such a standing wave takes place is defined as $H_i^\diamond < H_i < \omega/\mu_0\gamma$. The internal field $H_i$ increases with increase of a bias magnetic field $H_0$.

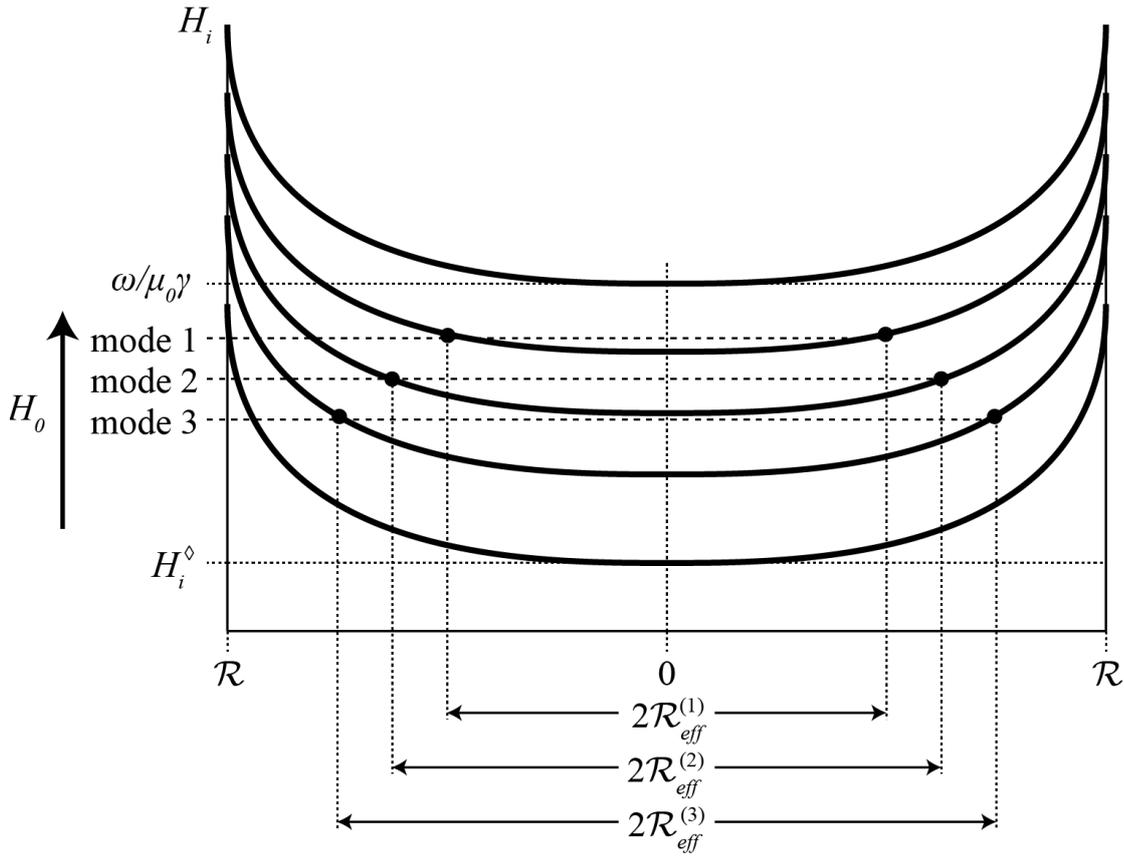

Fig. 10. A qualitative picture of the Bohr–Sommerfeld-quantization levels of an internal magnetic field $H_i$ for standing waves corresponding to the first three MDMs. With increasing the mode number, the effective diameter $2\mathcal{R}_{eff}^{(n)}$ increases as well.

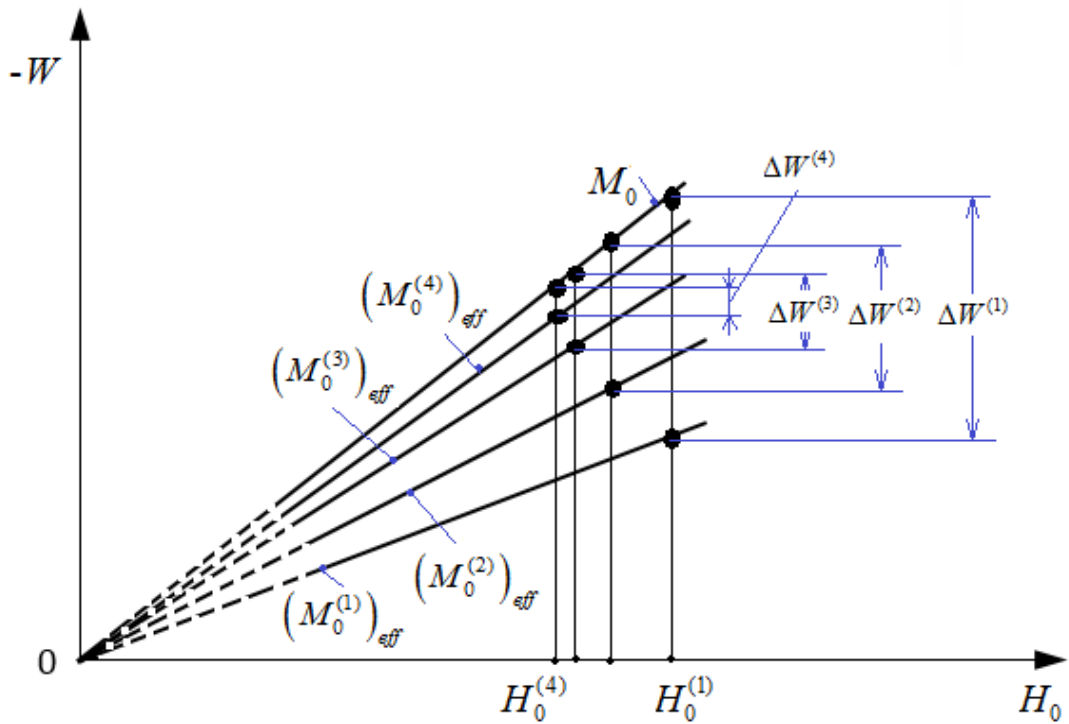

Fig. 11. A model illustrating discretization of magnetic energy in a ferrite disk at $\omega = const$. The slope of the straight lines is determined by DC magnetization. $M_0$ is saturation magnetization of a homogeneous ferrite material. At MDM resonances in a ferrite disk, the slope of the straight lines is determined by $\left(M_0^{(n)}\right)_{eff}$. $\Delta W^{(n)}$ shows the microwave energy extracted from a ferrite disk at the $n$-th MDM resonances. Discretization of the magnetic energy is shown for the first four MDMs.

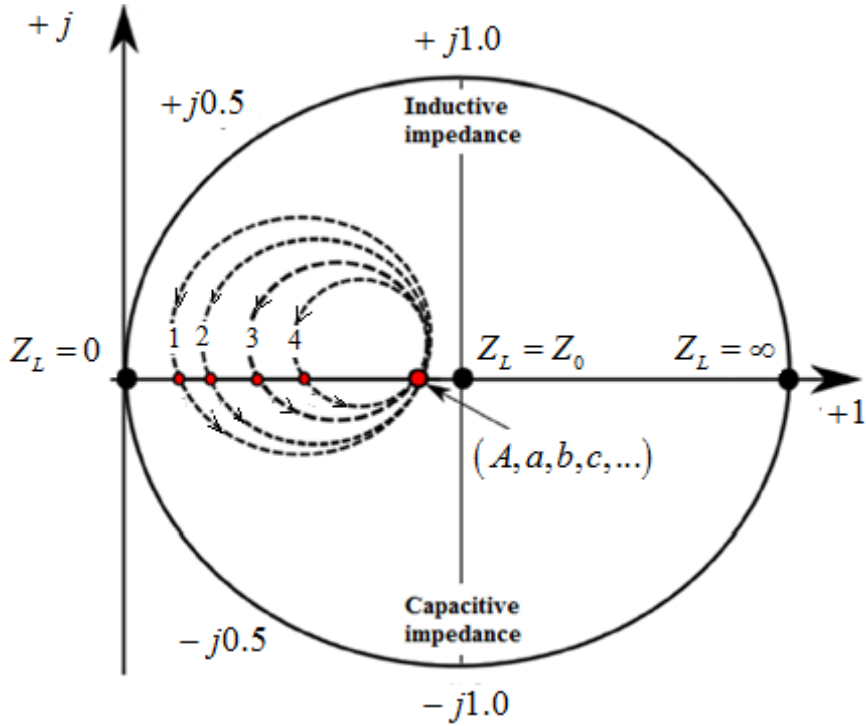

Fig. 12. Schematic representation of the cavity input impedances at MDM resonances on the complex-reflection-coefficient plane (the Smith chart). The numbers and letters correspond to numbers and letters in Fig. 8 (*b*). As the energy swept through an individual resonance, one observes evolution of the phase – the phase lapses. The phase jump of $\pi$ is observed each time a resonant condition is achieved.

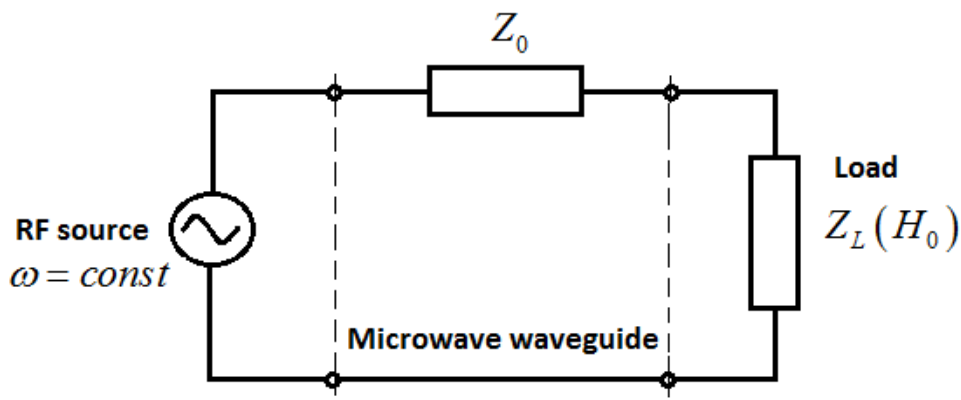

Fig. 13. An equivalent scheme describing a structure shown in Fig. 8. At a given frequency $\omega$, determined by a RF source, a microwave waveguide is presented by a characteristic impedance $Z_0$. The waveguide is loaded by an impedance $Z_L$, which depends on an external parameter – a bias magnetic field $H_0$.

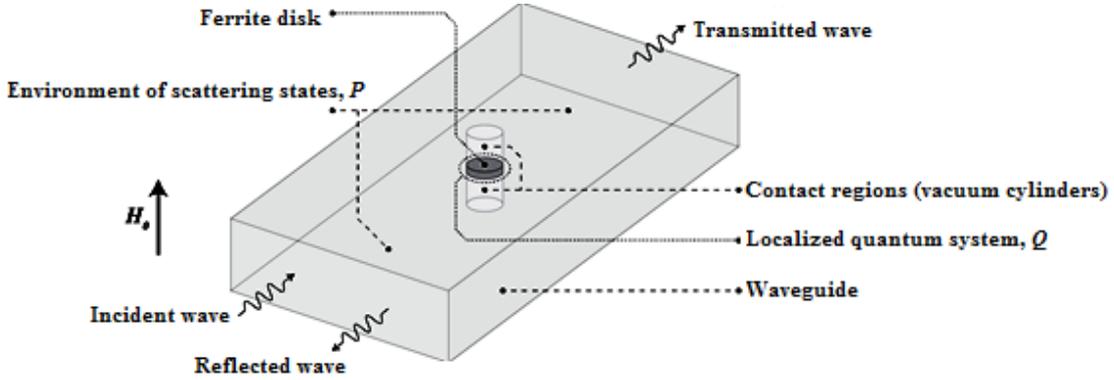

Fig. 14. An interaction of a MDM ferrite disk with a microwave waveguide. The structure is viewed as the $P + Q$ space. It consists of a localized quantum system (the MDM ferrite disk), denoted as the region $Q$, which is embedded within an environment of scattering states (the microwave waveguide), denoted as the regions $P$. The coupling between the regions $Q$ and $P$ is regulated by means of the two "contact regions" in the waveguide space.

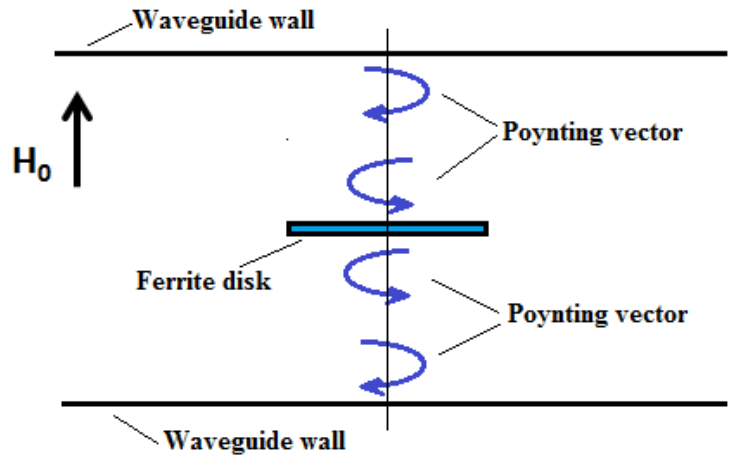

Fig. 15. Directions of rotations of the power-flow vortices along the axis of a contact region. In the contact regions, above and below a ferrite disk, we have helical-mode tunneling.

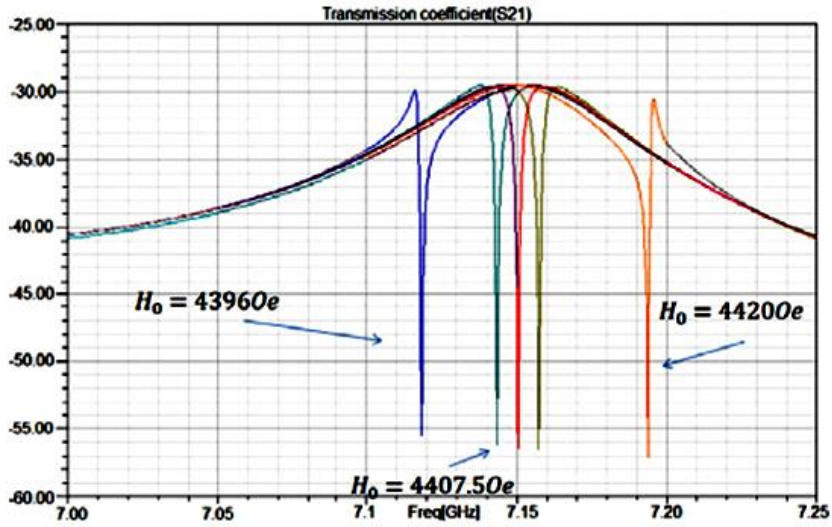

Fig 16. Modification of the Fano-resonance shape. At variation of a bias magnetic field, the Fano line shape of a MDM resonance can be completely damped. The scattering cross section of a single Lorentzian peak corresponds to a pure dark mode. Reference to [85].

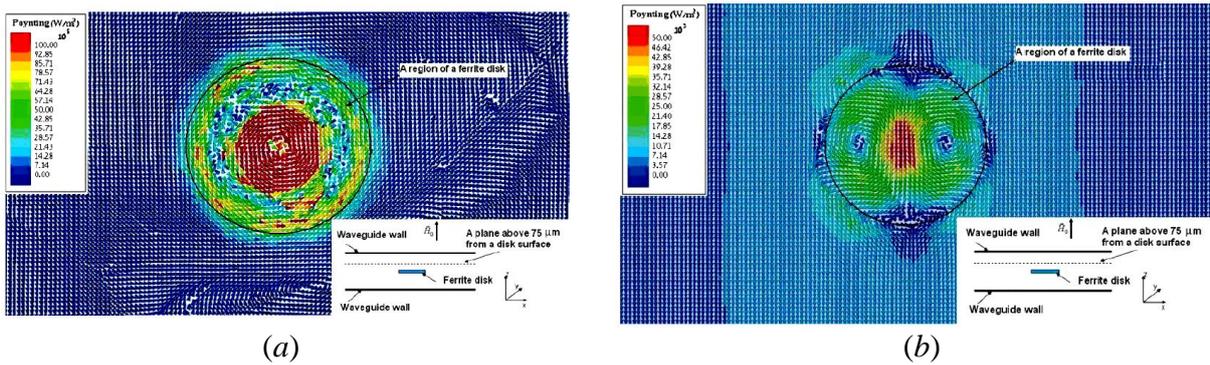

(a)          (b)

Fig. 17. An example of the power-flow density distributions in a near-field vacuum region above a ferrite disk. A ferrite disk is placed in a rectangular waveguide symmetrically to waveguide walls. (a) Single-rotating-magnetic-dipole (SRMD) resonance; (b) double-rotating-magnetic-dipole (DRMD) resonance. There is the dissimilar radiation rates of the SRMD and DRMD resonances. Radiation of the SRMD results in strong reflection while at the DRMD mode one has electromagnetic-field transparency and cloaking for a ferrite particle. Reference to [86].

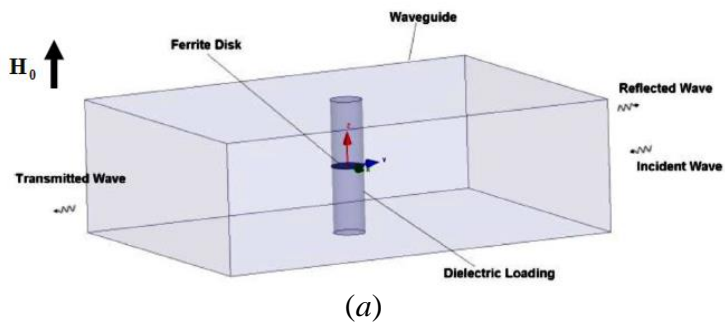

(a)

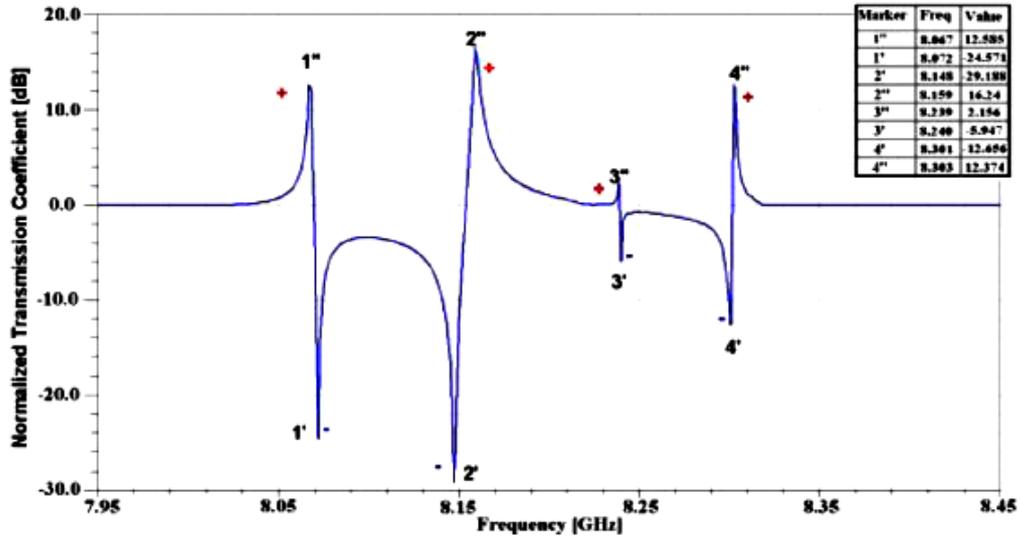
(b)

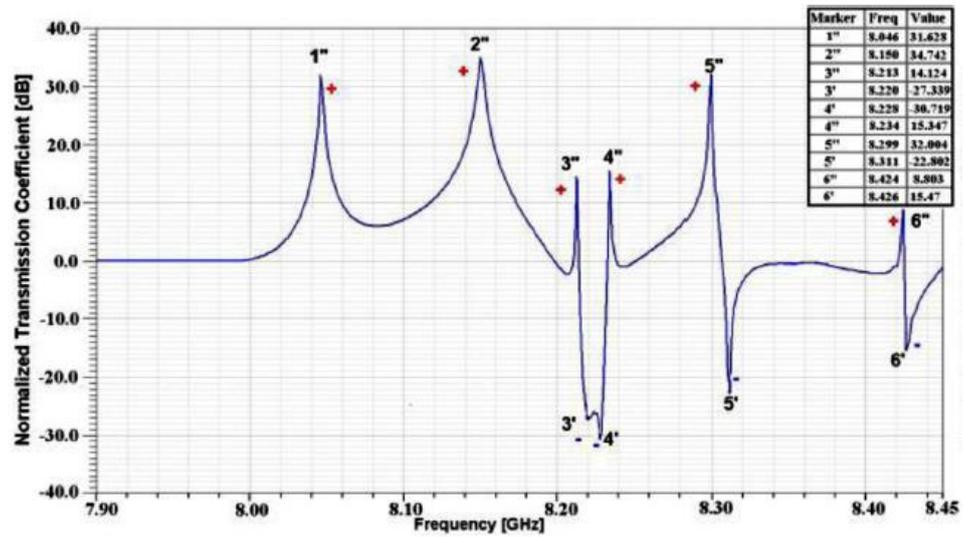
(c)

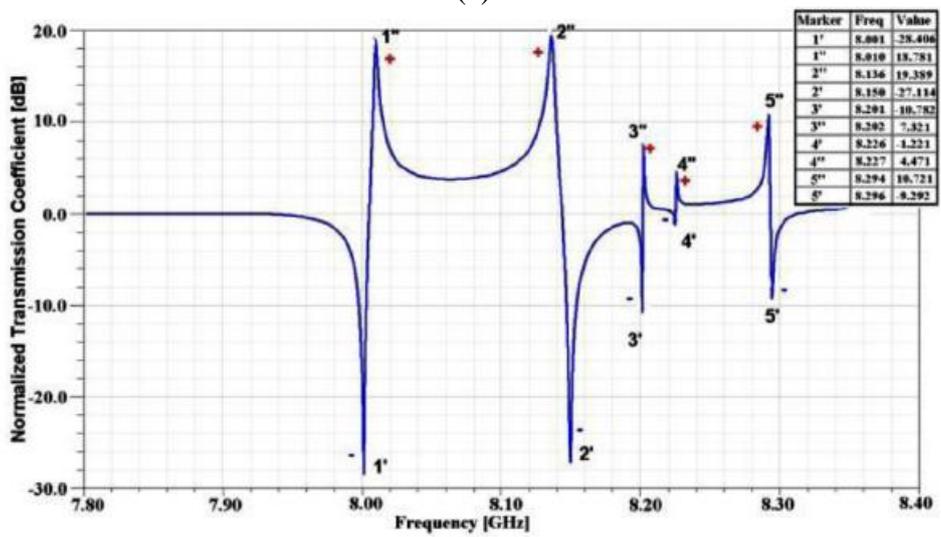
(d)

Fig. 18. Fano resonance collapse observed in the microwave transmission. (*a*) The structure of a rectangular waveguide with a MDM ferrite disk and loading dielectric cylinders. (*b*) Transmission characteristics at a dielectric parameter $\varepsilon_r = 30$, (*c*) at $\varepsilon_r = 38$, (*d*) at $\varepsilon_r = 50$. For the 1$^{st}$ and 2$^{nd}$ MDMs, the Fano line shapes are completely damped at $\varepsilon_r = 38$. In this case, single Lorentzian peaks appear. The scattering cross section corresponds to pure bright modes. Reference to [83].

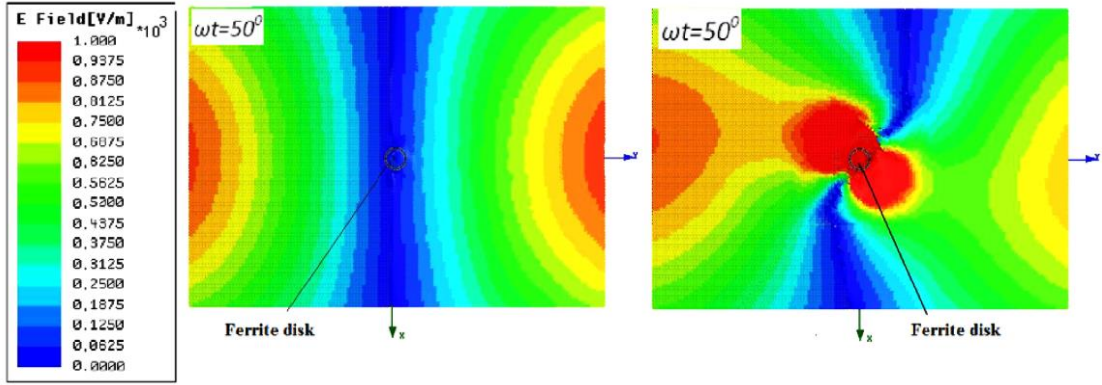

Fig. 19. Evidence for strong resonance field enhancement at the MDM resonance. Passing the front of the electromagnetic wave, when the frequency is (*a*) far from the frequency of the MDM resonance and (*b*) at the frequency of the MDM resonance in the disk. Reference to [82].

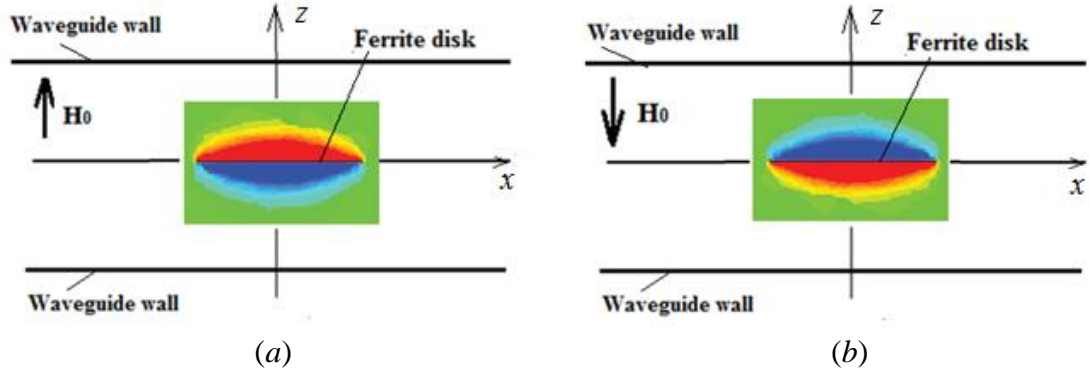

Fig. 20. The helicity factor $F^{(E)}$ distribution near a ferrite disk in a geometrically symmetrical structure. In a red region, $F^{(E)} > 0$, in a blue region, $F^{(E)} < 0$, in a green region, $F^{(E)} = 0$. The factor $F^{(E)}$ is characterized by antisymmetrical distribution with respect to *z* axis. Along with this, the helicity factor changes a sign at time reversal.

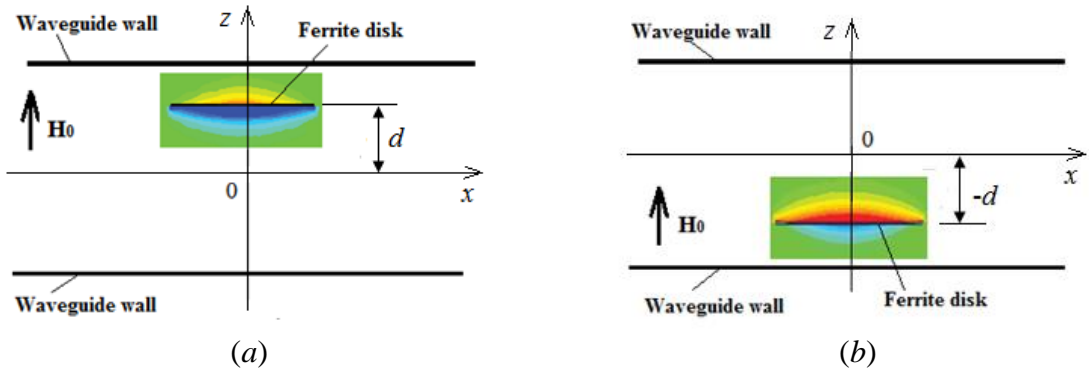

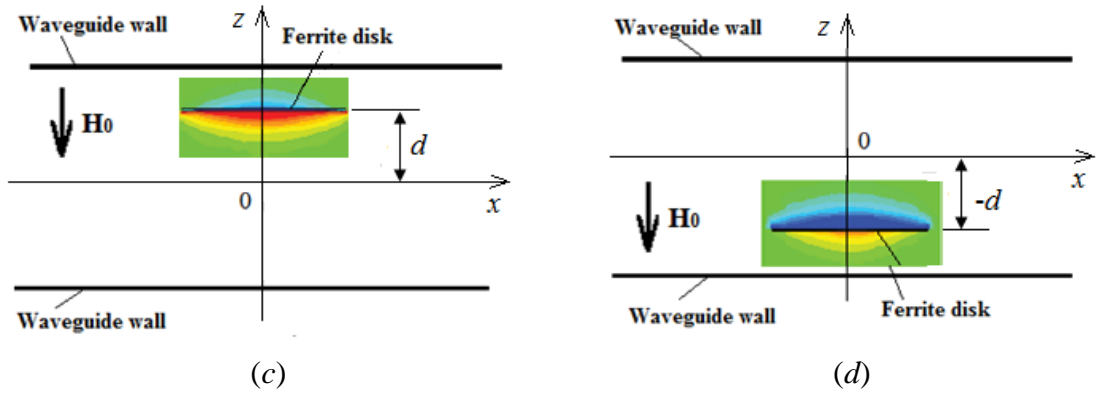

(c)      (d)

Fig. 21. Four cases of the helicity-factor distributions when two external parameters – the disk position on $z$ axis and the direction of a bias magnetic field – change. When a MDM ferrite disk is placed in a geometrically nonsymmetric structure, the distribution of a helicity factor $F^{(E)}$ becomes nonsymmetric as well. Reflection with respect to the $xy$ plane with simultaneous change of a direction of a bias magnetic field completely restore the same picture of a helicity factor $F^{(E)}$ [look at the distributions (a) and (d) and the distributions (b) and (c)]. At such a nonsymmetric disk position, the *PT* symmetry (with respect to the symmetry plane of the disk) of MDM oscillations is broken and the biorthogonality condition is not satisfied.

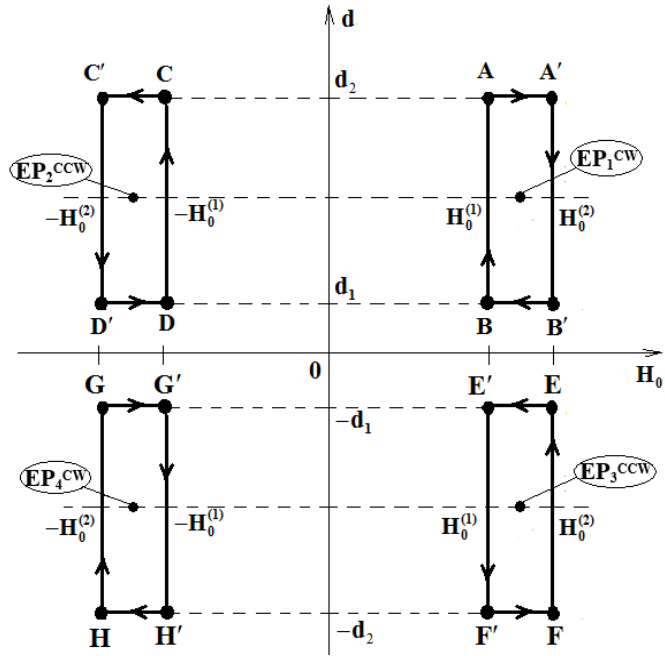

Fig. 22. Four types of contours encircling exceptional points along the paths on a 2D parametric space, a bias magnetic field and a disk shift.
Exceptional point $\mathbf{EP_1^{CW}}$ is inside the clockwise contour: $A \rightarrow A' \rightarrow B' \rightarrow B \rightarrow A$.
Exceptional point $\mathbf{EP_2^{CCW}}$ is inside the counterclockwise contour: $C \rightarrow C' \rightarrow D' \rightarrow D \rightarrow C$.
Exceptional point $\mathbf{EP_3^{CCW}}$ is inside the counterclockwise contour: $E \rightarrow E' \rightarrow F' \rightarrow F \rightarrow E$.
Exceptional point $\mathbf{EP_4^{CW}}$ is inside the clockwise contour: $G \rightarrow G' \rightarrow H' \rightarrow H \rightarrow G$.

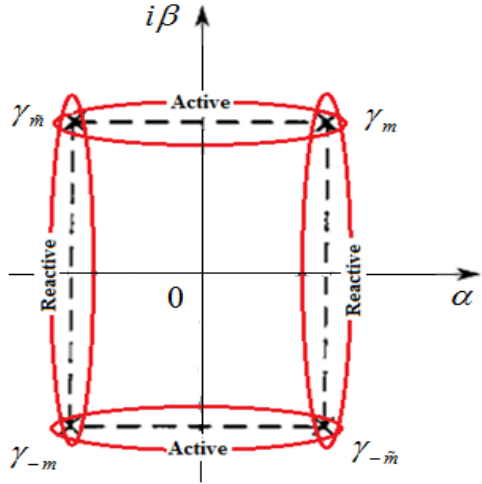

Fig. 23. Positions of the waveguide wave numbers on a complex plane.

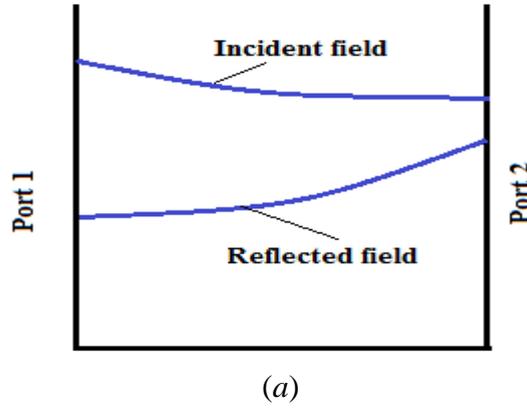

(*a*)

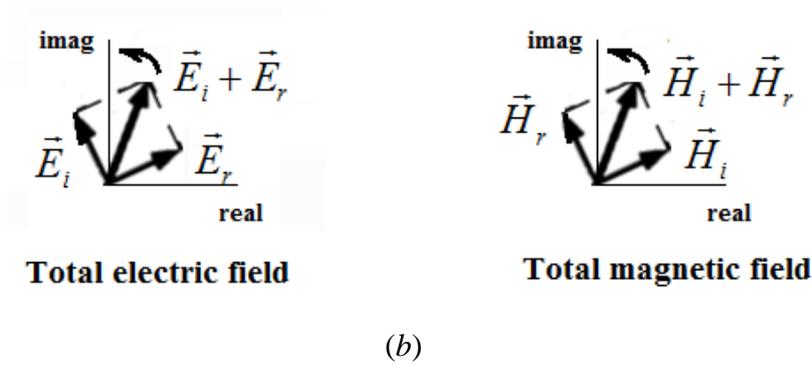

(*b*)

Fig. 24. The complex-plane incident and reflected field conditions for a below-cutoff waveguide section. (*a*) The incident (at port 1) field is the positive decaying mode and the reflected (from port 2) field is the negative decaying mode. (b) Due to the boundary conditions at port 2, the electric and magnetic fields of conjugate modes are 90° phase shifted in time.

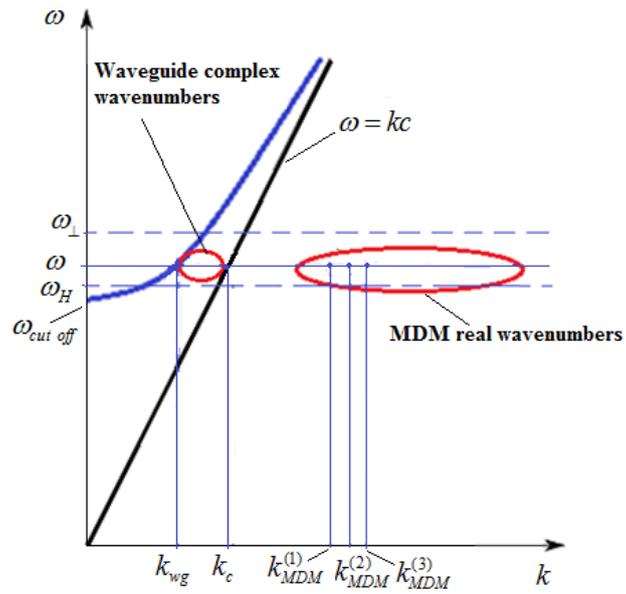

Fig. 25. Waveguide dispersion characteristics and MDM-oscillation BICs in a microwave waveguide. The MDM wavenumbers are beyond the region of the waveguide modes. The figure shows a region of existence of complex waveguide modes.